\documentclass[
  aps,
  prmaterials,
  reprint,
  amsmath,
  amssymb,
  superscriptaddress,
  floatfix
]{revtex4-2}
\usepackage{algorithm}
\usepackage{algpseudocode}
\usepackage{amsthm}
\usepackage{braket}
\usepackage{graphicx}
\usepackage{multirow}
\usepackage{pifont}
\usepackage{xcolor}
\usepackage[version=4]{mhchem}
\usepackage{tablefootnote}
\usepackage{float}
\usepackage{gensymb}
\usepackage{adjustbox}
\usepackage{xr-hyper}
\usepackage{hyperref}
\usepackage{dblfloatfix}

\begin{document}

\title{Large-scale first-principle simulations of amorphous indium oxide}
\author{Matthew Bousquet}
\email{bousquet@uchicago.edu}
\affiliation{Department of Chemistry, The University of Chicago, Chicago, Illinois 60637, USA}

\author{Francois Gygi}
\email{fgygi@ucdavis.edu}
\affiliation{Department of Computer Science, University of California Davis, Davis, CA 95616, USA}

\author{Giulia Galli}
\email{gagalli@uchicago.edu}
\affiliation{Pritzker School of Molecular Engineering, The University of Chicago, Chicago, Illinois 60637, USA}
\affiliation{Department of Chemistry, The University of Chicago, Chicago, Illinois 60637, USA}
\affiliation{Materials Science Division, Argonne National Laboratory, Lemont, Illinois 60439, USA}

\begin{abstract}
Amorphous indium oxide (a-In$_2$O$_3$) is a high-electron-mobility semiconductor of central importance in thin-film transistors and a promising photoanode for solar-driven water oxidation. Despite sustained experimental and computational investigations, the structural motifs underlying its unusual transport properties and the existence of O–O peroxide-like bonds within its network have remained unresolved. Here we develop a MACE-based machine-learned interatomic potential trained on first-principles molecular dynamics trajectories and use it to generate and analyze amorphous structures containing up to 5120 atoms, two orders of magnitude larger than those adopted in typical ab initio studies. We find X-ray structure factors in excellent quantitative agreement with experiment and we confirm that In$_2$O$_3$ is a poor glass former, with the likely presence of quasi-crystalline regions in amorphous samples. Our large-scale structural analysis reveals extended chains of edge-sharing InO$_k$ polyhedra providing a concrete structural basis for the high electron mobility of a-In$_2$O$_3$. Our results strongly support the formation of O–O peroxide-like bonds in the amorphous network,  with a mean length of $\sim$1.5 Å. We show that these bonds introduce localized in-gap states near the conduction band minimum, acting as a source of intrinsic n-type self-doping and enhancing sub-gap optical absorption. These effects are detectable via a distinct Raman feature near 850 cm$^{-1}$ that is absent in the IR spectrum. Overall, our results establish a comprehensive structure–property picture of a-In$_2$O$_3$, provide directly testable experimental predictions, and suggest that controlled amorphization is a viable strategy for improving the photoelectrochemical activity of a-In$_2$O$_3$.
\end{abstract}

\maketitle 

\section{Introduction} 

\typeout{XR-TEST: si.aux loaded = \@ifundefined{r@fig:somelabel}{NO}{YES}}

Amorphous indium oxide (a-In$_2$O$_3$) is a technologically relevant amorphous oxide semiconductor owing to its combination of high electron mobility, optical transparency, and compatibility with low-temperature deposition processes \cite{Charnas2023, Yap2023, Si2022}. These properties make it a leading channel material for thin-film transistors (TFTs) \cite{Charnas2023, Si2022, Kim2022, Yuvaraja2024, Seong2025, Hu2025} and a promising component in complementary metal-oxide-semiconductor (CMOS) architectures, where the ability to deposit uniform, high-quality films at modest temperatures greatly simplifies device fabrication compared with single-crystal approaches.
A defining and practically important characteristic of a-In$_2$O$_3$ is the retention of high, band-like electron mobility upon amorphization \cite{Buchholz2013, Buchholz2014, Jankousky2026}. This behavior is uncommon, as the loss of long-range translational order in amorphous materials usually leads to a substantial increase in carrier scattering and thereby to a degradation of electronic transport. In a-In$_2$O$_3$, an $s$ orbital character at the conduction band minimum (CBM) has been invoked to explain the persistence of high mobility, but the precise structural motifs responsible for the high mobility —including the role of edge- and corner-sharing polyhedra—have not yet been established.
Another largely unresolved question concerns the formation of O–O bonds in the a-In$_2$O$_3$ network.  Computational studies have debated both the existence and the electronic properties of such bonds \cite{Aliano2011, Medvedeva2020defect, Jankousky2026}, which have also been proposed to exist in other amorphous oxides \cite{Arhammar2011}.

Addressing the open questions highlighted above requires simulations of systems substantially larger than the 40-80 atom supercells that have dominated the computational literature to date \cite{Jankousky2026}. Small cells impose constraints on the sampling of disorder, the description of long-range network connectivity, and the statistical characterization of rare structural motifs such as O–O pairs. The use of  small cells in conjunction with first-principles molecular dynamics (FPMD) simulations has been motivated by the relatively high computational cost  of this technique for systems with many hundreds of atoms and for simulation times exceeding hundreds of picoseconds. Machine-learned interatomic potentials (MLIPs), which in principle can reproduce first-principles potential energy surfaces at a fraction of the computational cost, offer a practical path forward.

In this work, we develop and validate a MACE-based MLIP for indium oxide, trained on FPMD trajectories spanning liquid, crystalline, and amorphous configurations. The equivariant message-passing architecture of MACE and the availability of a foundation model make it attractive for fine-tuning on oxides. The MLIP is used to generate ensembles of a-In$_2$O$_3$ structures at densities ranging from 6.92 to 7.19 g cm$^{-3}$ in supercells of up to 5120 atoms—two orders of magnitude larger than typical FPMD studies reported to date. We characterize the structural properties through radial distribution functions (RDFs), angular distribution functions, and polyhedral connectivity analysis, benchmarking against experimental X-ray scattering data. Vibrational properties are analyzed through the calculation of Raman and IR spectra. Electronic structure calculations using both  SCAN\cite{Sun2015} and a dielectric-dependent hybrid (DDH)\cite{Skone2014} functional are performed on representative snapshots to investigate band-edge character and charge localization. We also elucidate the optical activity of amorphous indium oxide and whether it can be improved relative to that of the crystal and hence a-In$_2$O$_3$ may be a promising water splitting photo-electrode.  Our previous work \cite{Bousquet2025} showed that the bands of crystalline In$_2$O$_3$ are well positioned to catalyze the two-electron water oxidation reaction (WOR). However, the valence band maximum (VBM) to conduction band minimum (CBM) transition is optically dark, and strategies to lower the optical gap are required. Amorphization of the material is a possible strategy to decrease the onset of optical activity and facilitate the WOR.

The remainder of the paper is organized as follows. Section 2 describes the computational methods, including our DFT calculations, melt-and-quench protocol, MLIP training procedure, and analysis methods. Section 3 presents the structural and vibrational properties of a-In$_2$O$_3$ and section 4 presents the electronic structure results. Section 5 summarizes our findings and outlines future directions.

\section{Computational Methods}

\subsection {First-Principles Molecular Dynamics and Electronic Structure Calculations}

All FPMD simulations were performed using the Qbox code \cite{Gygi2008} with the SCAN meta-generalized gradient approximation (meta-GGA) exchange-correlation functional \cite{Sun2015} used in our previous work to study crystalline photoelectrodes \cite{Bousquet2025}. Optimized norm-conserving Vanderbilt (ONCV) pseudopotentials \cite{Schlipf2015} were employed with a plane-wave kinetic energy cutoff of 65 Ry. Brillouin zone integration was restricted to the $\Gamma$ point. Equations of motion were integrated with a time step of 40 a.u. ($\simeq$ 0.96 fs) and temperature was maintained at 330 K using the Bussi-Donadio-Parrinello \cite{Bussi2007} (BDP) thermostat. 

Amorphous structures were generated using a melt-and-quench procedure. In our FPMD simulations, an 80-atom cubic supercell of In$_2$O$_3$ was heated at a rate of 140 K ps$^{-1}$ and equilibrated at 5000 K for approximately 8 ps to yield a diffusive liquid. Thirty-one independent trajectories --23 at 7.19 g $cm^{-3}$, 4 at 7.11 g $cm^{-3}$, and 4 at 6.92 g $cm^{-3}$--were then branched and cooled to 330 K at constant rates spanning 100–450 K ps$^{-1}$. These densities were selected for their relevance to the use of amorphous indium oxide in thin-film transistors. Of the 31 trajectories, 27 were used to train a machine-learned potential, as explained in detail in the SI.

Electronic structure calculations on selected snapshots extracted from FPMD and MLIP trajectories were carried out using both SCAN and the dielectric-dependent hybrid (DDH) exchange-correlation functional \cite{Skone2014}. In this approach, the fraction of exact (Hartree-Fock) exchange, $\alpha$, is determined as the inverse of the macroscopic electronic dielectric constant: ${\alpha = {\epsilon_{\infty}}^{-1}}$, which was evaluated at the hybrid SE-RSH level of theory\cite{Zhan2025} for crystalline indium oxide, yielding $\epsilon_{\infty} = 4.2$. The experimental values reported in the literature vary between 4.0–4.1 for the crystal \cite{Feneberg2016} and between  3.3–4.0 for the amorphous, depending  on film density,  and residual crystallinity~\cite{Sundaresh2021}.

\subsection{Training Set Generation and Model Training}

We trained a MLIP starting from the MACE foundation model\cite{batatia2022mace, batatia2025crosslearning} and using configurations extracted from FMPD trajectories as our initial training set. We then carried out active learning cycles first on 80 atom and then on 640 atom supercells. The final MLIP was used to carry out MD simulations with cells up to 5120 atoms. Details of the chosen training and validation sets and of the potential generation procedure are given in the SI.

DFT training-data calculations for the MLIP were performed with Quantum ESPRESSO \cite{giannozzi2009qe, Giannozzi2020} with a plane-wave cutoff of 90 Ry and a charge-density cutoff of 1200 Ry.
We emphasize that a large charge density cutoff is critical and it was required for the following reasons. In Neural Network architectures translational invariance is implicitly enforced via construction of geometric features from relative interatomic distance vectors ($\mathbf{r}_{ji} = \mathbf{r}_j - \mathbf{r}_i$ )in order to preserve the symmetry of the underlying potential energy surface. In DFT calculations with plane-wave basis sets, the total energy is exactly invariant under translations at integer multiples of the grid vectors, whose spacing is determined by the plane-wave (PW) cutoff chosen to represent the charge density (usually 4 times as that of the wavefunction). We present an example in the SI for silicon, where we achieve an excellent translational invariance with moderate energy cutoff when using the LDA functional\cite{Perdew1981lda, Ceperley1980, Kohn1965} and, to some extent the PBE functional \cite{perdew1996pbe}. However a much larger charge density cutoff (and hence coarser grid spacing) is required for the SCAN functional which contains a complex expression of the kinetic energy. In In$_2$O$_3$, errors of up to 1 eV/\AA{} in forces were found with a cutoff of 260 Ry for the charge density (corresponding to a plane-wave cutoff of 65 Ry), when using SCAN.

\subsection{Calculation of radial distribution functions and structure factors}
Partial radial distribution functions (RDF) $g_{\alpha\beta}(r)$ for In–In, In–O, and O–O pairs were computed adopting the common definition:
\begin{equation}
g_{\alpha\beta}(r) = \frac{V}{N_\alpha N_\beta}\left\langle \sum_{i \in \alpha} \sum_{j \in \beta}\frac {\delta(r - r_{ij})}{4\pi r^2} \right\rangle
\label{eq:pgr}
\end{equation}
where $N_{\alpha (\beta)}$ is the number of atoms of species $\alpha$, ($\beta$) and $V$ is the simulation cell volume. 
A total radial distribution function was then constructed from the stoichiometric weighted sum of partial correlation functions: $
g_{\text{tot}}(r) = c_{\text{In}}^2\, g_{\text{InIn}}(r) + c_{\text{O}}^2\, g_{\text{OO}}(r) + 2\, c_{\text{In}} c_{\text{O}}\, g_{\text{InO}}(r)$,
with $c_{\text{In}} = 2/5$ and $c_{\text{O}} = 3/5$.

X-ray atomic form factors $f_\alpha$ used to define the structure factor were evaluated using the Cromer-Mann parameterization\cite{cromer1968}:
$
f_\alpha(\sin\theta / \lambda) = c_\alpha + \sum_{i=1}^{4} a_{\alpha,i}\, \exp\!\left[-b_{\alpha,i}\left(\sin\theta / \lambda\right)^{\!2}\right]$,
where $\lambda$ is the wavelength of the incident X-ray, $\theta$ is the scattering angle, and a, b and c are the Cromer--Mann coefficients taken from the International Tables for Crystallography\cite{Brown2006}. We used ionic coefficients for both In and O.
Partial structure factors were computed directly and also  via Fourier transformation of the partial RDFs:
\begin{equation}
S_{\alpha\beta}(k) = 1 + \frac{4\pi \rho_0}{k} \int_0^{r_{\max}} r\,\bigl[g_{\alpha\beta}(r) - 1\bigr]\, \sin(kr)\, dr,
\end{equation}
where $\rho_0 = (N_{\text{In}} + N_{\text{O}}) / V$ is the total number density. The total reduced X-ray interference function $i(k)$ was obtained from the combination of the partial structure factors \cite{faber1965}:
\begin{widetext}
\begin{equation}
i(k) = \frac{c_{\text{In}}^2 f_{\text{In}}^2 (S_{\text{InIn}} - 1) + c_{\text{O}}^2 f_{\text{O}}^2 (S_{\text{OO}} - 1) + 2 c_{\text{In}} c_{\text{O}} f_{\text{In}} f_{\text{O}} (S_{\text{InO}} - 1)}{\langle f \rangle^2},
\label{eq:ik_gr}
\end{equation}
\end{widetext}

where $c_\alpha = N_\alpha / N_{\text{tot}}$ are the atomic fractions and $\langle f \rangle = c_{\text{In}} f_{\text{In}}(k) + c_{\text{O}} f_{\text{O}}(k)$ is the mean form factor. 

When computing the coherent intensity directly in reciprocal space, 
for each frame, the partial density was computed as:
$
\rho_\alpha(\mathbf{k}) = \sum_{j \in \alpha} \exp(-i\,\mathbf{k}\cdot\mathbf{r}_j)$,
with the total scattering amplitude being
$A(\mathbf{k}) = f_{\text{In}}(k)\, \rho_{\text{In}}(\mathbf{k}) + f_{\text{O}}(k)\, \rho_{\text{O}}(\mathbf{k})$
and the coherent intensity 
$I_{\text{coh}}(\mathbf{k}) = \bigl\langle\, \lvert A(\mathbf{k}) \rvert^{2} \,\bigr\rangle.$
Intensities were spherically averaged and the reduced interference function was defined as:
\begin{equation}
i(k) = \frac{I_{\text{coh}}(k)/N_{\text{tot}} - \langle f^2 \rangle}{\langle f^2 \rangle},
\qquad
\langle f^2 \rangle = c_{\text{In}} f_{\text{In}}^2(k) + c_{\text{O}} f_{\text{O}}^2(k),
\label{eq:ik_direct}
\end{equation}

\subsection{ Calculations of Vibrational Properties}
The vibrational density of states (VDOS) was computed from the Fourier transform of the velocity autocorrelation function extracted from SCAN NVT trajectories for 80- and 640-atom cells, both for samples with and without O–O bonds. For computational simplicity, Raman spectra were computed at the LDA level for 80-atom cells using the finite-displacement method to evaluate the dynamical matrix and Raman tensors, as implemented in the Quantum Espresso Code.\cite{Lazzeri2003raman} We note that our calculations of the VDOS with SCAN and LDA yield similar and consistent results.

\section{Structural and Vibrational Properties}
\subsection{Radial distribution functions and structure factors}

\begin{figure*}
  \centering
  \includegraphics[width=1\linewidth]{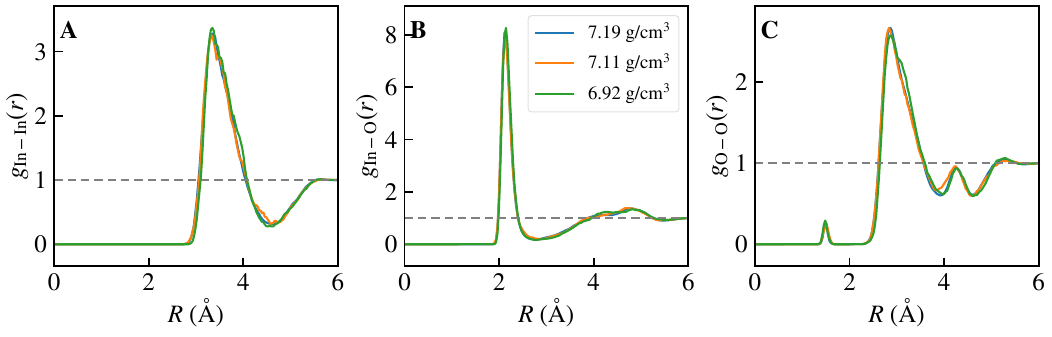}
  \caption{Partial radial distribution functions, g$_{In-In}$(r) (A), g$_{In-O}$(r) (B), and g$_{  O-O }$(r) (C), computed with first principle molecular dynamics for 80 atom amorphous samples as a function of density. Standard deviations are reported in Figure \ref{fig:si_pgr_rho} and coordination numbers in Table S1, showing that In-In coordination decreases to 5.48-5.42 from the value of 6 in the crystal, and In-O coordination to 3.65-3.61, from the value of 4 in the crystal, depending on the density.}
  \label{fig:partials}
\end{figure*}

We begin by comparing our FPMD results for a-In$_2$O$_3$, obtained with 80 atom cells,  at three densities: 7.19, 7.11, and 6.92 g cm$^{-3}$. As seen in Figure \ref{fig:partials} the radial distribution functions are remarkably similar over the density range considered here. We note the presence of a  peak in the  O-O  partial $g(r)$ at 1.5 \AA{}, corresponding to an  O-O  bond, further discussed in detail in section \ref{section:oo}.  

\begin{figure}[tb]
  \centering
  \includegraphics[width=1\linewidth]{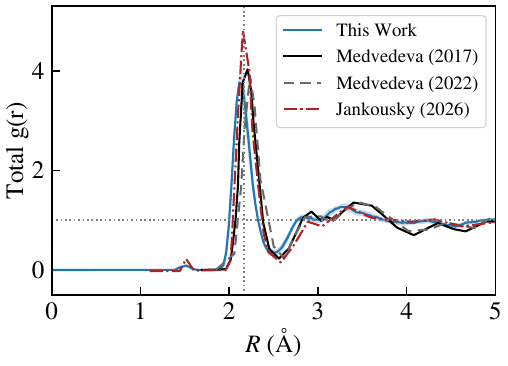}
  \caption{Comparison of the computed total correlation function with previous results, reported in references \cite{Medvedeva2017, Medvedeva2022, Jankousky2026} at $\rho$=7.11 g cm$^{-3}$.}
  \label{fig:tgr_lit}
\end{figure}

The total $g(r)$ is reported in Figure \ref{fig:tgr_lit}, where it is compared with those reported in  the literature (see also Figure \ref{fig:sio_lit_pgr} in the SI). We find an overall qualitative agreement with previous results, with some quantitative discrepancies on the position of the first peak, corresponding to In-O distances: our computed average In-O bond distance is 2.12 $\pm$ 0.02 \AA{}, versus 2.20 \AA{} reported by Medvedeva et al.\cite{Medvedeva2017,Medvedeva2022},  and 2.17 \AA{} by Khanal et al. \cite{Khanal2015}. All previous works adopted the PBE functional\cite{perdew1996pbe}, instead of SCAN and the use of different functionals is likely responsible for the difference in the computed distances. Quantitative discrepancies are also observed in the position of the second double peak of the total $g(r)$.

\begin{figure*}
  \centering
  \includegraphics[width=1\linewidth]{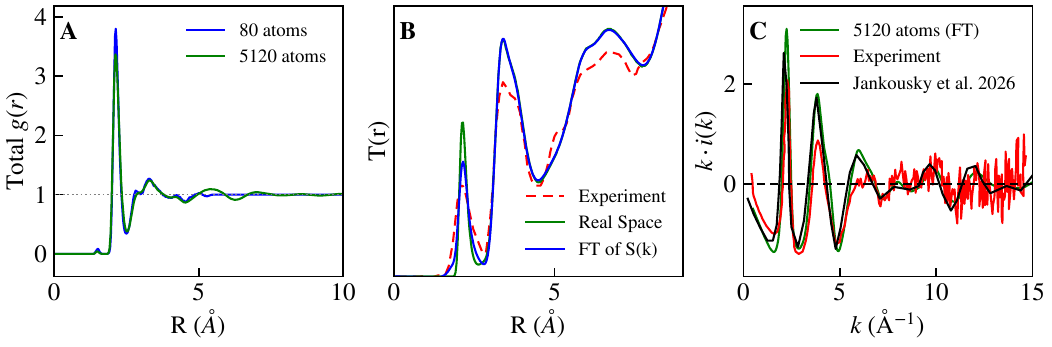}
  \caption{\textbf{A:} Total radial distribution function ($g(r)$) of a-In$_2$O$_3$ computed with 80 and 5120 supercells. \textbf{B:} Comparison between experimental \cite{Utsuno2006} and computed correlation function $(T(r) = 4\pi\rho r \cdot g(r)$) obtained directly in real space and as a Fourier Transform (FT) of the structure factor S (k)\textbf{C:} Comparison between measured and computed interference functions (see Eq. 4) Results from Ref.\cite{Jankousky2026} are also shown. 
  }
  \label{fig:3panel_struc_v1}
\end{figure*}

 In Figure \ref{fig:3panel_struc_v1}A  we compare the total radial distribution function $g(r)$ of a-In$_2$O$_3$ computed with 80- and 5120-atom supercells, the latter results obtained with the ML potential. We find nearly identical $g(r)$ profiles in the range 1.0 - 5.0 \AA{}, demonstrating that the short- and intermediate-range orders in a-In$_2$O$_3$ are well-captured even by relatively small simulation cells. However, the data obtained with the 5120 atom supercell show additional peaks between 5 and 10 \AA{}, indicating the presence of a third coordination shell and of a less pronounced fourth coordination shell.

Figure \ref{fig:3panel_struc_v1}B and C show a comparison of our data with experiment. In B, we report the function
$T(r) = 4\pi\rho r \cdot g(r)$
 computed  directly in real space from the total $g(r)$ and  as the Fourier transform (FT) of the X-ray-weighted reduced interference function k$\cdot$i(k), mimicking the procedure adopted to obtain radial distribution functions from experimental structure factors. 
 In C we show the reduced interference function k$\cdot$i(k) computed with  5120 atom cells compared with X-ray data and published results. We find an excellent agreement with experiments and also good agreement with previous works, with minor quantitative discrepancies on the peak positions, as in the case of the total $g(r)$. 

\subsection{Peroxide oxygen bonds}
\label{section:oo}
A subtle but electronically significant structural motif in a-In$_2$O$_3$ is the formation of short O–O bonds with bond length close to 1.5 \AA{}, reminiscent of peroxide linkages. The existence of such bonds has been debated in the literature \cite{Aliano2011, Medvedeva2020defect, Jankousky2026}, partly because small simulation cells (e.g., 40 atoms) may insufficiently sample the configurational space required to observe or rule out events leading to  O-O  bonds formation. As mentioned in the introduction, the presence of O–O bonds has been reported in other amorphous oxides \cite{Arhammar2011}. 

\begin{figure}
  \centering
  \includegraphics[width=1\linewidth]{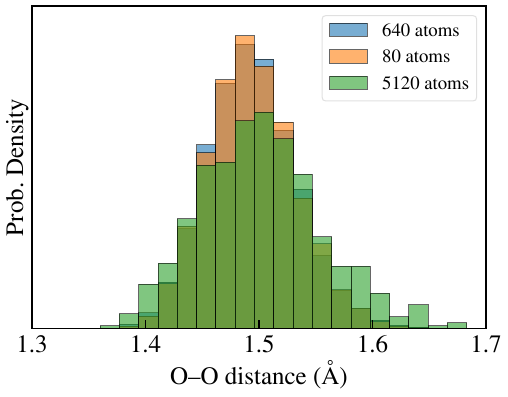}
  \caption{Histogram of  O-O bond lengths for samples generated by our MLIP at $\rho$ = 7.11 g $\cdot$ cm $^{-3}$. The mean distance is 1.49 $\pm$ 0.04 \AA{} for 80 and 640 atom cells and 1.50 $\pm$ 0.05 \AA{} for the 5120 cell.}
  \label{fig:oo-hist}
\end{figure}

Our results strongly support the presence of O-O bonds in amorphous indium oxide at the densities considered here.  In the 80 atom samples, we observe O-O bonds in 50$\%$ of the generated samples, whether used in training (51 samples) or production runs (50 samples). In the 640 atom samples, 40 out of 45 samples present O-O bonds (e.g. almost 90$\%$ of samples), with a majority of them (34) having 3 to 5 O-O bonds and the rest exhibiting between 1 and 2. Finally, we generated one 5120 sample that has 32 O-O bonds, amounting to 2$\%$ of oxygen atoms in the system forming such bonds, with an average first neighbor distance between their center of mass of 1 $\pm$ 0.12 nm. We find that for all sample sizes, the average distance of O-O bonds is 1.5 \AA, with a sizable spread between 1.4 and 1.6 \AA (see Figure \ref{fig:oo-hist}). 

\begin{figure}
  \centering
  \includegraphics[width=1\linewidth]{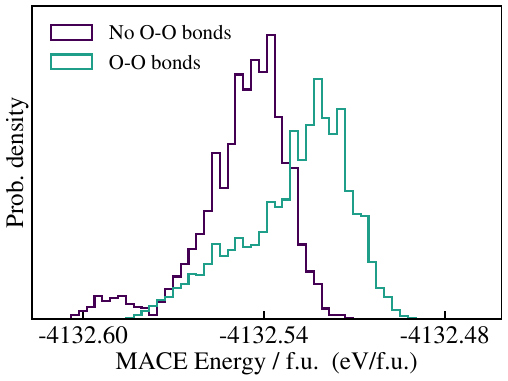}
  \caption{Distribution of MLIP energies for 640-atom systems, with a without O–O bonds. Although the two distributions overlap considerably, samples without O–O bonds tend to exhibit lower energies on average. 
  Note the small peak at low energy, where we find quasi-crystalline structures in the absence of O-O bonds.}
  \label{fig:640-e-hist}
\end{figure}

A detailed analysis of the trajectories obtained for the 640 atom samples show that configurations without O-O bonds are on average lower in energy relative to those with O-O bonds, however the energy differences (see Figure \ref{fig:640-e-hist}) are relatively small, on the order of 0.015 eV/formula unit, and consistent with the differences found in the 80 atom samples. 

During one of our 640 atom melt and quench simulations with the MLIP we recovered a fully crystalline structure that contained a peroxide interstitial defect. From this structure we computed a defect formation energy of 0.04 $\pm$ 0.007 eV per formula unit (average on 10 snapshots), only 3 times higher than that found for the amorphous. These results suggest that, not surprisingly, O-O  bonds can be better stabilized in an amorphous than in a crystalline environment. 
 
Interestingly, in the generated 640 atom trajectories with and without O-O bonds we could identify a broad distribution of configurations, some of which with higher order of crystallinity and with lower energy, as shown in Figure \ref{fig:640-e-hist}. These results suggest that depending on the preparation conditions, a-In$_2$O$_3$ oxide  samples may present nearly crystalline regions embedded in an amorphous matrix, consistent with the fact that indium oxide is found experimentally to be a poor glass forming material. We find that on average the difference in energy per formula unit between amorphous and fully crystalline samples is 0.7 $\pm$ 0.05 eV/f.u., not dissimilar to that measured in other poor glass forming solids, e.g. zirconia, for which an energy difference of 0.6 eV/f.u. was measured by calorimetry \cite{Ellsworth1994, Molodetsky2000}  and alumina, for which a difference of 0.9 eV/f.u. was reported from ab-initio simulations \cite{Aykol2018}.



 
To characterize the vibrational properties of a-In$_2$O$_3$, we computed the vibrational density of states (VDOS) at the SCAN level of theory for structures with and without O–O bonds, using both 80-atom and 640-atom supercells. The general features of the VDOS are consistent between cell sizes: we find a broad distribution of modes spanning 0–700 cm$^{-1}$, broadly assignable to In–O stretching, bending, and libration modes of the InO$_k$ polyhedra.

\begin{figure}
  \centering
  \includegraphics[width=1\linewidth]{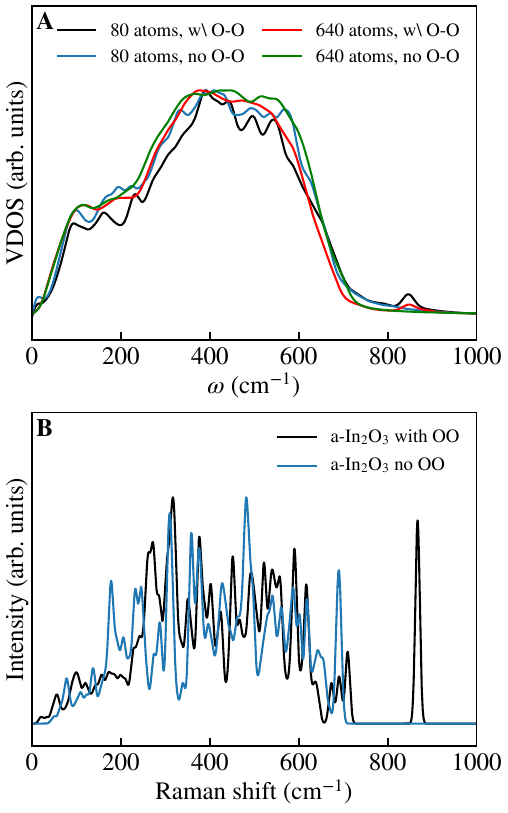}
  \caption{Panel A shows the vibrational density of states (VDOS) for 80 atom samples of a-In$_2$O$_3$ with and without an  O-O bond and the VDOS of a 640 atom sample with 4 O-O bonds and one without 0 O-O bonds. The peak around $\sim$ 850 cm$^{-1}$ is attributed to the short ($\sim$ 1.50 \AA{}) peroxide bond. Panel B shows Raman spectra computed at the LDA level of theory for a 80 atom cell with and without an O-O  bond. The mode near $\sim$ 850 cm$^{-1}$ is Raman-active, and corresponds to the  O-O  bond}
  \label{fig:aIO_vib}
\end{figure}

A distinctive feature of the VDOS in structures containing O–O bonds is the appearance of a well-defined peak near 850 cm$^{-1}$ (Figure \ref{fig:aIO_vib}). This feature is absent in O–O-free structures and it arises from the stretching vibration of the short O–O peroxide-like bond. An analogous mode appears near the same frequency in hydrogen peroxide, providing a useful benchmark for the assignment. The sharpness and position of this feature are found in calculations for all cell sizes, demonstrating that it is an intrinsic signature of the O–O bond.

To assess experimental detectability, we computed the Raman spectrum for an 80-atom cell containing an O–O bond, and we used the LDA functional for simplicity (Figure \ref{fig:aIO_vib}). The computed Raman spectrum reproduces the broad envelope of In–O modes in the 200–700 cm$^{-1}$ range in good agreement with available measurements of In$_2$O$_3$ \cite{Kranert2014raman, Stokey2021raman}. Importantly, the O–O stretch near 850 cm$^{-1}$ appears as a distinct Raman-active feature, demonstrating that Raman spectroscopy could serve as a direct experimental probe of O–O bond concentration in amorphous films. Notably, this feature is absent from the IR spectra, shown in figure \ref{fig:aIO_ir}.

\subsection {Angular Distribution Functions and Polyhedral Connectivity}

To characterize the intermediate-range of a-In$_2$O$_3$, we computed the In–O–In bond-angle distributions for corner-sharing and edge-sharing InO$_k$ polyhedral configurations (Figure \ref{fig:aIO_adf_rho}). Corner-sharing configurations, in which two polyhedra share a single bridging oxygen, exhibit a broad distribution centered near 120–130\degree. Edge-sharing configurations, in which two polyhedra share a common edge (two bridging oxygens), display a narrower distribution centered near 95–100\degree. The ratio of edge to corner shared polyhedra is similar for all densities, as seen in Figure \ref{fig:poly_frac}. Standard deviations are provided in section \ref{section:adf_error} of the SI.

\begin{figure}[tb]
  \centering
  \includegraphics[width=1\linewidth]{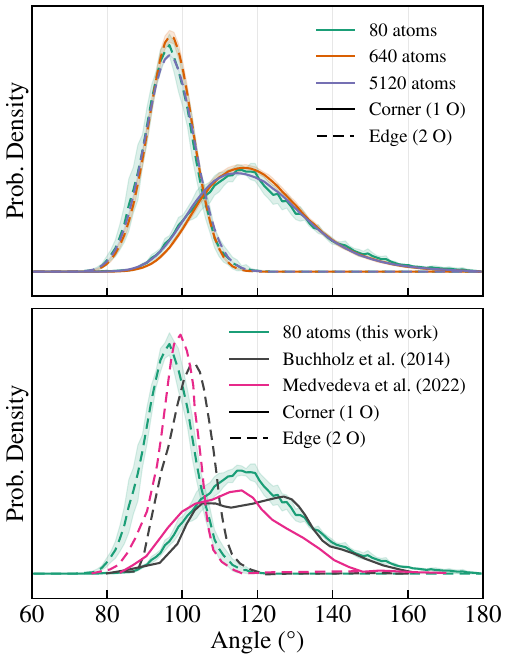}
  \caption{ Computed angular distribution functions (upper panel) and comparison of  first-principles molecular dynamics results with published angular distribution functions, reported by   Buchholz et al\cite{Buchholz2014} and Medvedeva et al.\cite{Medvedeva2022} (lower panel).}
  \label{fig:adf_lit}
\end{figure}

 Interestingly, as seen in Figure \ref{fig:adf_lit}, the angular distribution functions are similar when obtained with 80, 640 and 5120 atom supercells. Our results are in reasonable agreement with those obtained in previous works and reported by Buchholz et al\cite{Buchholz2014} and Medvedeva et al (2022) \cite{Medvedeva2022}, with some notable differences. The most significant one emerges in the edge-sharing configurations (dashed lines), where our FPMD results yield a peak centered at approximately 97°, which is notably lower than the 104° reported by Buchholz et al. and closer to the 99° found by Medvedeva et al. The shape of the edge-sharing distributions also differ : Buchholz et al. show a broad distribution shifted to higher angles, while Medvedeva et al. display a sharper, more well-defined peak. 

\begin{figure}[tb]
  \centering
  \includegraphics[width=1\linewidth]{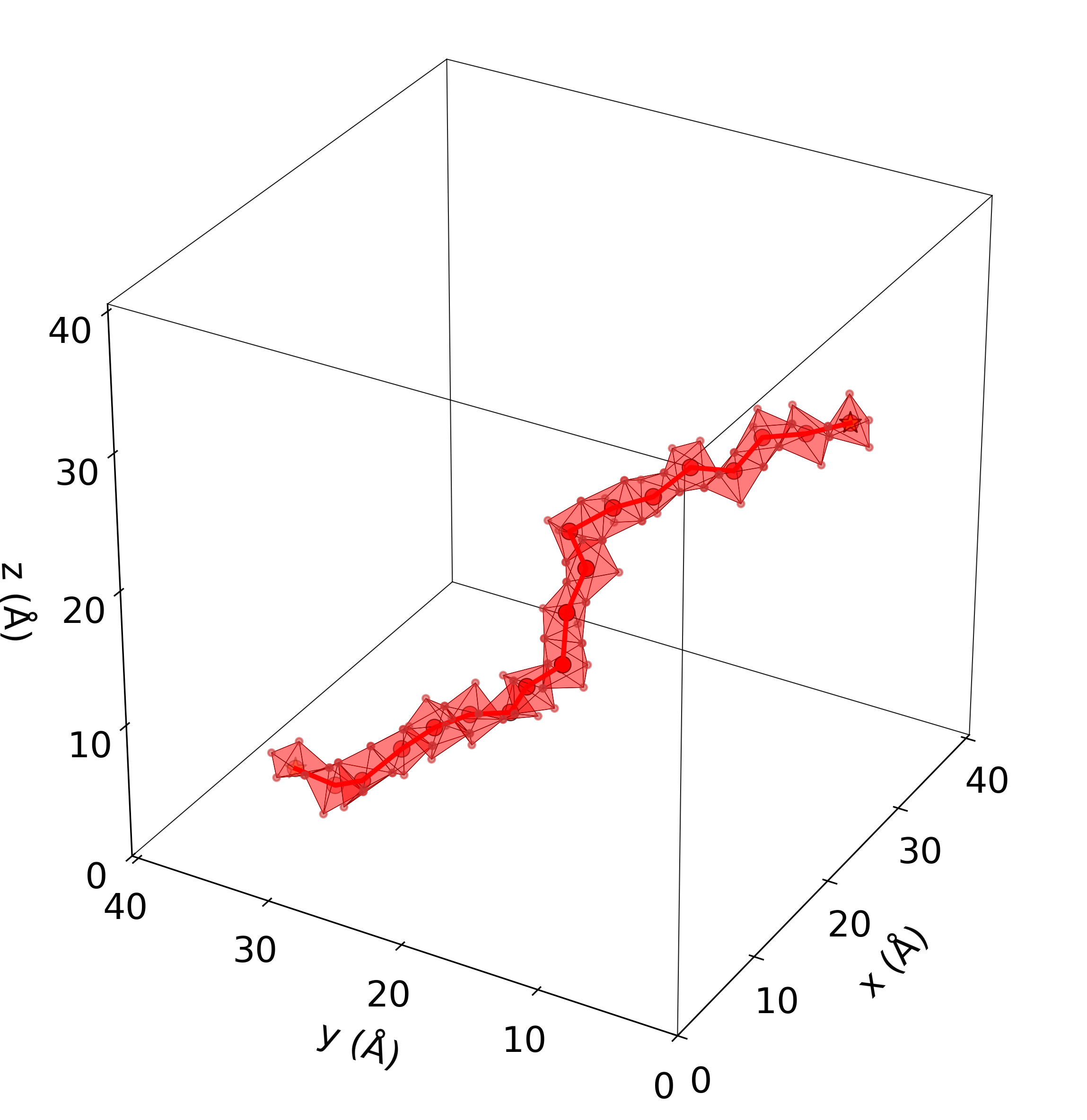}
  \caption{Longest chain of InO edge sharing polyhedra detected in a  5120 atom supercell with a density of 7.11 g cm$^{-3}$.}
  \label{fig:5120_chain}
\end{figure}

In our large cells with 5120, we identified the presence of long-range chains of edge and face connected InO$_k$ polyhedra. The average chain length (see Figure \ref{fig:5120_chain_stats}) is 12.18 $\pm$ 2.41 In atoms, spanning 20.82 $\pm$ 6.25 \AA{}, with some chains extending across the entire cell (see Fig. \ref{fig:5120_chain}) In addition we find that only approximately 2$\%$ of polyhedra in the system are disconnected from others, showing the persistence of a percolating network of polyhedra functioning as potential  pathways for electron conduction. The identification of such network, which cannot be made with small supercells, represents an important result of our study. 



\section{Electronic Structure}
\label{section:elec}
\begin{figure}[tb]
  \centering
  \includegraphics[width=1\linewidth]{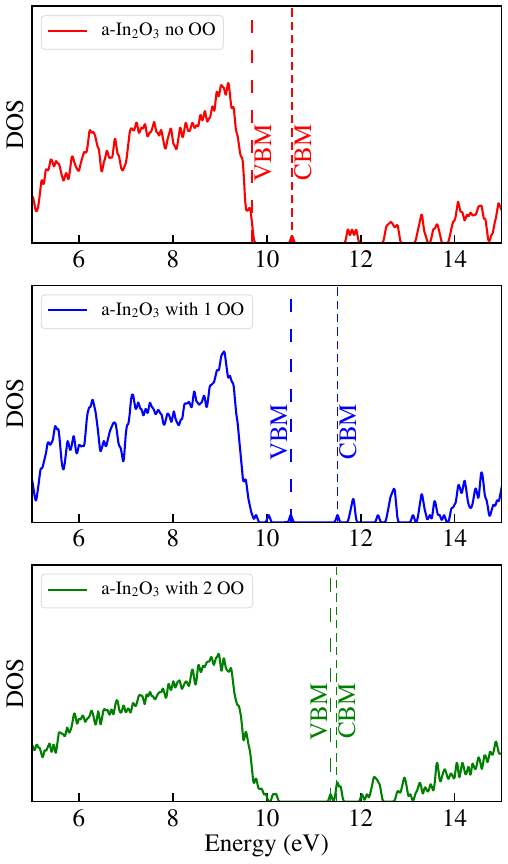}
  \caption{Electronic Density of States (EDOS) for a-In$_2$O$_3$ with and without an  O-O bond. The top, middle and bottom panel show results for a 640 atom cell containing no O-O bonds, 1  O-O  and 2  O-O bonds, respectively. On each plot, the valance band maximum (VBM) and conduction band minimum (CBM) are shown. The introduction of  O-O bonds pushes the highest occupied orbitals close to the bottom of the conduction band.}
  \label{fig:aIO_edos_v3}
\end{figure}

\begin{figure}
  \centering
  \includegraphics[width=1\linewidth]{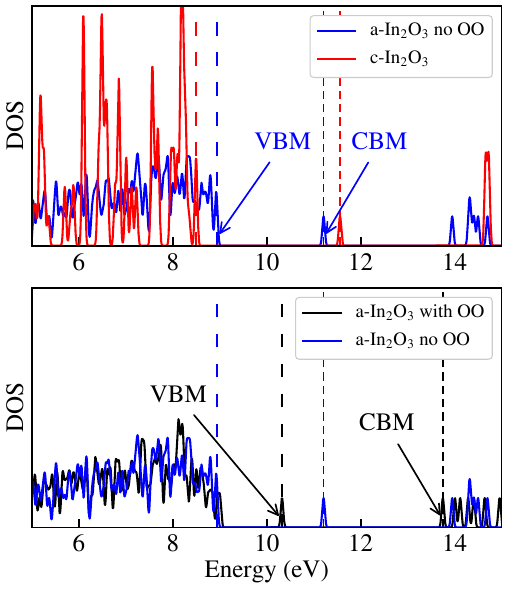}
  \caption{Electronic Density of States (EDOS) computed for 80 atom cells at the hybrid (DDH level) level of theory.  We show three samples: c-In$_2$O$_3$, a-In$_2$O$_3$ without an  O-O  bond and a-In$_2$O$_3$ with an O-O bond. At the hybrid level, the gap of the amorphous system shrinks to 2.27 eV from 3.06 eV in the crystal. Note the difference in the density of conduction states in 80 and 640 atom samples (see Fig. \ref{fig:aIO_edos_v3}) due to finite size effects. A Gaussian broadening of 0.025 eV was applied to all energy levels.}
  \label{fig:80_cIO_hybrid}
\end{figure}

Having established the structural properties of a- In$_2$O$_3$, we turn to investigate its electronic structure.  The electronic density of states (EDOS) of structures computed at the SCAN level of theory for representative snapshots from the MLIP trajectories with 640 atoms, with and without O-O bonds, is shown in (Figure \ref{fig:aIO_edos_v3}).  The top panel presents the case without O-O bonds and shows a wide gap between O 2p orbitals and In 5s orbitals, similar to the crystal. Consistent with the literature \cite{Medvedeva2020defect, Aliano2011, Medvedeva2017}, we find that the valence band maximum (VBM) is broadened relative to the crystalline phase, with Urbach tail states extending into the gap. These tail states are associated with under-coordinated or strained In–O bonds and act to reduce the gap of a- In$_2$O$_3$ compared with c- In$_2$O$_3$. We also computed the electronic structure of an amorphous sample with the hybrid functional DDH, shown in Figure \ref{fig:80_cIO_hybrid}, however with 80 atom cells, given the cost of using hybrid functionals for 640 atom cells.  We found results consistent with our SCAN results, although the numerical values of the band gaps are different, as expected.  The band gap  at the DDH level of theory is 2.27 eV, a reduction of 0.79 eV relative to that of c-In$_2$O$_3$. The VBM shifts upward by 0.43 eV, relative to the crystal, corresponding to the presence of the tail states mentioned above. Such a shift is beneficial for the use of a-In$_2$O$_3$ as a photoanode for water oxidation, as it pushes the valence band edge closer to the water oxidation redox potentials. 

\begin{figure}
  \centering
  \includegraphics[width=1\linewidth]{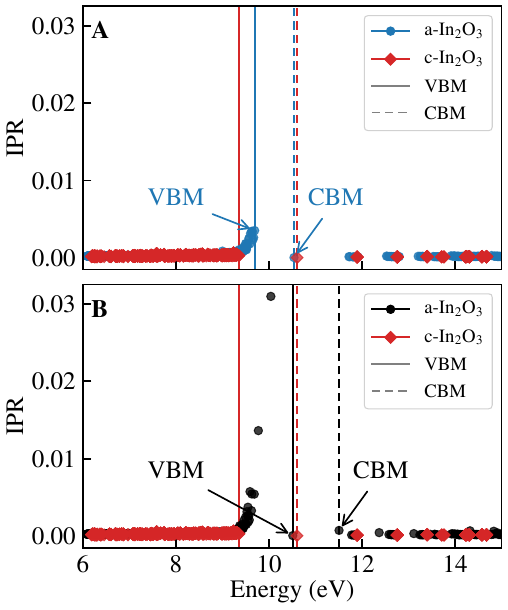}
  \caption{Inverse Participation Ratio (IPR) of the energy levels  of a-In$_2$O$_3$ with and without an  O-O bond, compared to that computed for the crystal, in a 640 atom cell. In the absence of the  O-O bond, local valence band tail states lead to a reduction in the gap. When an O-O bond is present, we observe in-gap states highly localized on the peroxide bond and a highly delocalized occupied state, close to the conduction band minimum.}
  \label{fig:ipr}
\end{figure}

The inverse participation ratio (IPR) as a function of Kohn-Sham energy (Figure \ref{fig:ipr}, left) reveals that valence band tail states are more localized than bulk valence edges, while conduction edges remain essentially delocalized. This finding supports the view of band-like electron transport in a-In$_2$O$_3$, and is consistent with the experimentally observed high electron mobility.

The middle and bottom panels of Figure \ref{fig:aIO_edos_v3}  show the EDOS of samples with O-O bond, showing that such bonds introduce filled  states into the gap region. The highest occupied orbital, a delocalized state associated with a O-O bond, is close to the conduction band minimum, acting as a source of intrinsic n-type doping. This self-doping mechanism could contribute to the finite conductivity observed in nominally undoped a- In$_2$O$_3$ films and may explain some of the sample-to-sample variability in transport measurements, which may arise from a different concentration of O-O bonds in the sample.




\subsection{Optical Properties}

\begin{figure}[tb]
  \centering
  \includegraphics[width=1\linewidth]{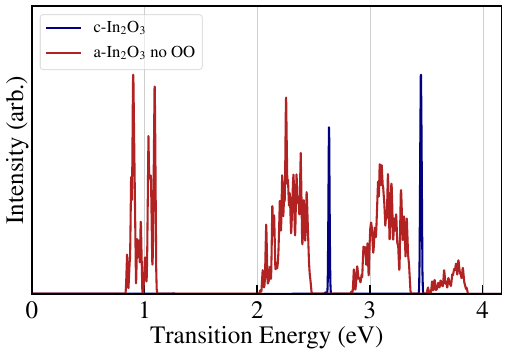}
  \caption{Dipole transition strengths from occupied orbitals to unoccupied orbitals for 640 atom cells. Notably, transitions from the valence band edge to the conduction band are optically active in a-In$_2$O$_3$, but dark in c-In$_2$O$_3$}
  \label{fig:tdms}
\end{figure}

Finally, we investigated the optical activity of a-In$_2$O$_3$ and we computed dipole transition strengths between occupied and unoccupied Kohn-Sham orbitals for 640-atom cells (Figure \ref{fig:tdms}), using the SCAN functional. In the absence of O–O bonds, the lowest-energy optical transitions correspond to excitations from VBM tail states to the CBM; these transitions are relatively weak and the onset of significant optical activity is determined by the bulk gap. The presence of O–O bonds introduces low-energy in-gap states that participate in optical transitions at energies below the main gap onset, potentially reducing the effective optical gap and increasing sub-gap absorption.
This enhanced sub-gap optical activity in structures with O–O bonds is relevant to the application of a-In$_2$O$_3$ as a photoanode for water oxidation reactions (WOR). Our prior work on c-In$_2$O$_3$ established that the band positions of In$_2$O$_3$ are thermodynamically favorable for WOR, but the fundamental gap transition is optically dark, limiting solar light absorption. Amorphization and the associated generation of O–O bonds offer a pathway to increase sub-gap optical activity and potentially improve the solar-to-hydrogen efficiency.

\section{Conclusions and Outlook}
We have presented a comprehensive computational study of the structural, vibrational, and electronic properties of amorphous In$_2$O$_3$ using a combination of first-principles molecular dynamics and a newly developed MACE-based machine-learned interatomic potential. By accessing supercells of up to 5120 atoms—far exceeding the scale of prior first-principles studies—we achieve quantitative agreement with experimental X-ray scattering data and establish statistically reliable structure–property relationships across mass densities from 6.92 to 7.19 g cm$^{-3}$.

We find that the local In–O coordination geometry of a-In$_2$O$_3$ is remarkably insensitive to density variations in this range, consistent with the experimentally observed robustness of electron mobility across differently deposited films. More significantly, our large supercells reveal for the first time extended chains of edge-sharing InO$_k$ polyhedra spanning the simulation cell. These percolating pathways, inaccessible in small supercells, provide a concrete structural rationale for high mobility enabled by band-like electron transport in a-In$_2$O$_3$.

A second central result is the identification and characterization of O–O peroxide-like bonds in the amorphous network. These bonds, with a mean length of 1.5 \AA{} and formation energies only modestly above those of O–O configurations in the crystal, are present in most of the independently generated 640-atom samples and in our generated 5120 sample. Electronic structure calculations show that the O–O bonds introduce in-gap states close to the conduction band minimum, which act as a source of intrinsic n-type doping and enhance sub-gap optical absorption. Inverse participation ratio analysis establishes both the spatial localization of these in-gap states and the delocalized character of the associated donor orbital. The vibrational signature of the O–O bond—a sharp mode near 850 cm$^{-1}$ present in the Raman spectrum but absent in the infrared—provides a direct experimental probe for quantifying O–O bond concentration in deposited films, and offers a practical handle for correlating deposition conditions with transport variability.

From a photochemistry perspective, amorphization lowers the optical gap by $\sim$0.8 eV relative to crystalline In$_2$O$_3$ and activates transitions that are dark in the crystal, making a-In$_2$O$_3$ a more promising photoanode candidate for the two-electron water oxidation reaction. Control over O–O bond population through deposition conditions thus emerges as a lever for simultaneously tuning the n-type carrier density, the sub-gap optical response, and potentially the photoelectrochemical efficiency.

Several directions merit future investigation. The interface of a-In$_2$O$_3$ with common gate dielectrics (SiO$_2$, HfO$_2$) involves structural relaxation, charge redistribution, and vacancy formation that critically influence device performance; the MLIP developed here provides the foundation for large-scale interface simulations at this level of accuracy. The role of O–O bonds and interfacial dipoles in neuromorphic device architectures, including ferroelectric field-effect transistors, also warrants systematic study. Finally, extending the present approach to explicitly model a-In$_2$O$_3$ surface and near-surface structures in contact with electrolyte will be essential for a first-principles description of the water oxidation mechanism.

\section{Acknowledgements}
This work was supported as part of the Advanced Materials for Energy-Water Systems Center, an Energy Frontier Research Center funded by the US Department of Energy, Office of Science, Basic Energy Sciences. The development of the MLIP based on FPMD data was supported by the computational materials science center MICCoM. We acknowledge the use of the computing resources provided by the University of Chicago Research Computing center and by NERSC.

\clearpage
\bibliography{bib-am}

\end{document}


\author{Matthew Bousquet}
\email{bousquet@uchicago.edu}
\affiliation{Department of Chemistry, The University of Chicago, Chicago, Illinois 60637, USA}

\author{Francois Gygi}
\email{fgygi@ucdavis.edu}
\affiliation{Department of Computer Science, University of California Davis, Davis, CA 95616, USA}

\author{Giulia Galli}
\email{gagalli@uchicago.edu}
\affiliation{Pritzker School of Molecular Engineering, The University of Chicago, Chicago, Illinois 60637, USA}
\affiliation{Department of Chemistry, The University of Chicago, Chicago, Illinois 60637, USA}
\affiliation{Materials Science Division, Argonne National Laboratory, Lemont, Illinois 60439, USA}

\title{Supporting Information for: Large scale simulations of amorphous indium oxide from first principles}

\maketitle
\setupSI

\tableofcontents
\newpage  

\section{Translation Invariance in DFT calculations and the derivation of MLIPs}

We show below convergence tests of energy and forces as a function of plane wave cutoff for crystalline Si and amorphous indium oxide. 

\subsection{Translation invariance in calculations for crystalline Si as a function of the exchange-correlation functional}

\begin{figure}[!ht]
    \centering
    \includegraphics[width=1\linewidth]{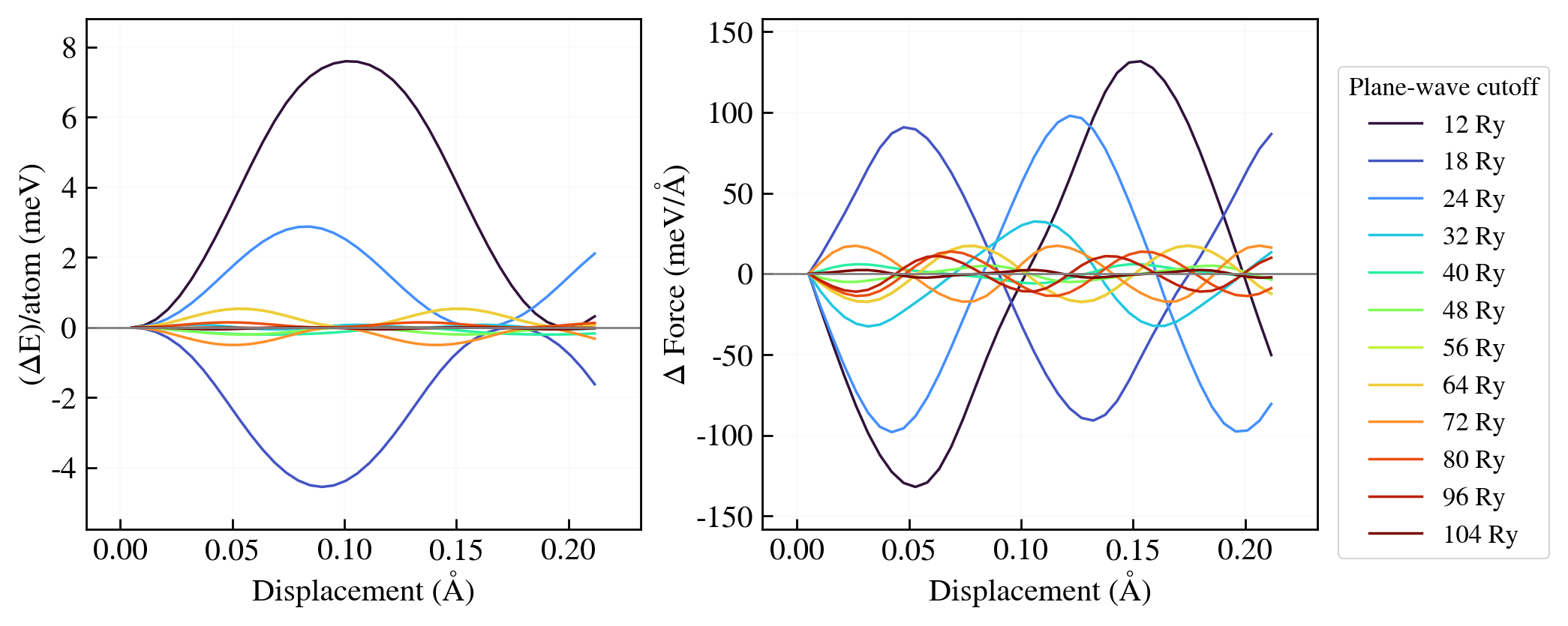}
    \caption{Total energies (left panel) and forces (right panel) computed at the DFT level of theory with the SCAN {\bf ref}) functional, as a function of ionic displacement, for different values of the plane wave cutoff, for a 64 atom Si cell. A  plane-wave cutoff as high as 104 Ry is required to  reduce the fluctuations of forces to those found at the PBE level of theory.}
    \label{fig:si_si_scan}
\end{figure}

Fig. S1 shows that when using a kinetic energy cutoff of $E_{\text{cut}} = 12$ Ry, with the local density approximation (LDA) \cite{Perdew1981lda} we obtain a negligible force amplitude of 0.05 meV/\AA{} over a displacement of 0.2 A, while the Perdew-Burke-Ernzerhof (PBE)\cite{perdew1996pbe} functional results in a slightly larger amplitude of 7.74 meV/\AA{}. The SCAN\cite{Sun2015} functional shows a significantly higher force amplitude of 132 meV/\AA{}. Increasing $E_{\text{cut}}$ to $104$ Ry successfully reduces the force amplitude obtained with SCAN to that of PBE, with a max amplitude of 2.4 meV/\AA{}, as shown in \ref{fig:si_si_scan}. 

\begin{figure}[!ht]
    \centering
    \includegraphics[width=1\linewidth]{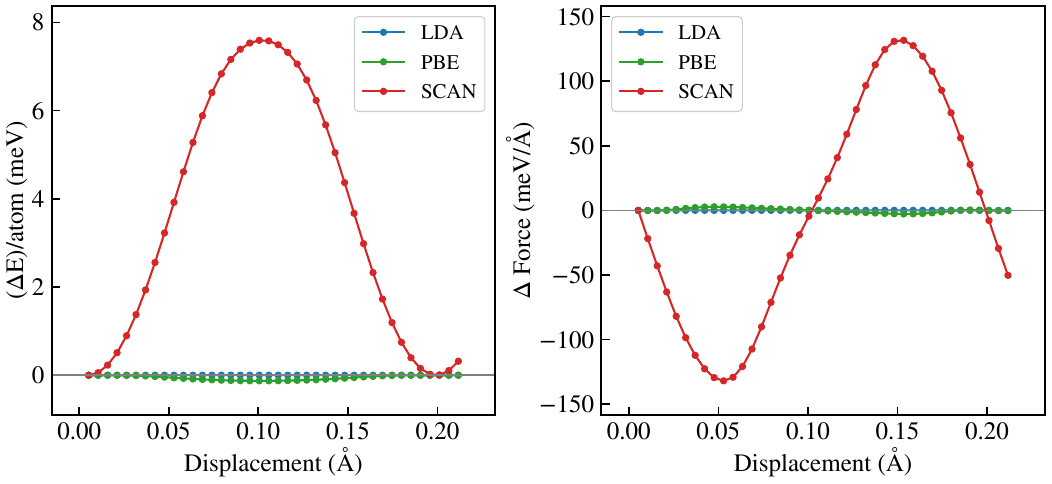}
    \caption{Total energies (left panel) and forces (right panel) computed at the DFT level of theory with three different functionals (LDA \cite{Perdew1981lda}, PBE \cite{perdew1996pbe} and SCAN \cite{Sun2015}), as a function of ionic displacement, with a plane wave cutoff of 12 Ry for a 64 atom Si cell. Forces computed with SCAN show a variation three orders of magnitude greater than those obtained with LDA or PBE.}
    \label{fig:si_functionals}
\end{figure}

\subsection{Translational invariance in calculations for a-In$_2$O$_3$}

We tested the convergence of total energy and forces as a function of planewave cutoff when using the SCAN functional. We note that while the total energy converges at about 65 Ry, the error in forces is much harder to converge and requires a cutoff as high as 1200 Ry.

\begin{figure}[!ht]
    \centering
    \includegraphics[width=1\linewidth]{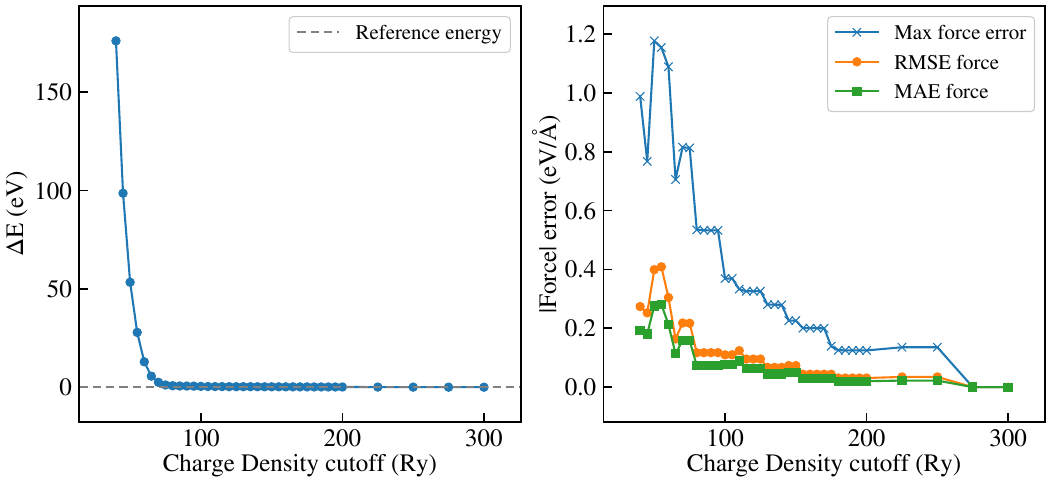}
    \caption{Convergence of the total energy (left) and force errors (right) with respect to the plane wave cutoff, computed with the SCAN functional and ONCV pseudopotentials for a single amorphous In$_2$O$_3$ (a-In$_2$O$_3$) snapshot; the charge density cutoff is held at four times the plane-wave cutoff. All errors are taken relative to the highest-cutoff calculation, which serves as the reference for both the energy and the forces. The force panel reports three measures of the per-atom force error $\Delta_i = \lvert \mathbf{F}_i - \mathbf{F}_i^{\mathrm{ref}} \rvert$: the maximum error $\max_i \Delta_i$, the root-mean-square error $\mathrm{RMSE} = \sqrt{\tfrac{1}{N}\sum_{i=1}^{N} \Delta_i^{2}}$, and the mean absolute error $\mathrm{MAE} = \tfrac{1}{N}\sum_{i=1}^{N} \Delta_i$, where the sum runs over all $N$ atoms. Although the total energy converges rapidly, the maximum force error remains appreciable even after the energy is well converged.}
    \label{fig:si_aIO_forces_noecutrho}
\end{figure}

\begin{figure}[!ht]
    \centering
    \includegraphics[width=1\linewidth]{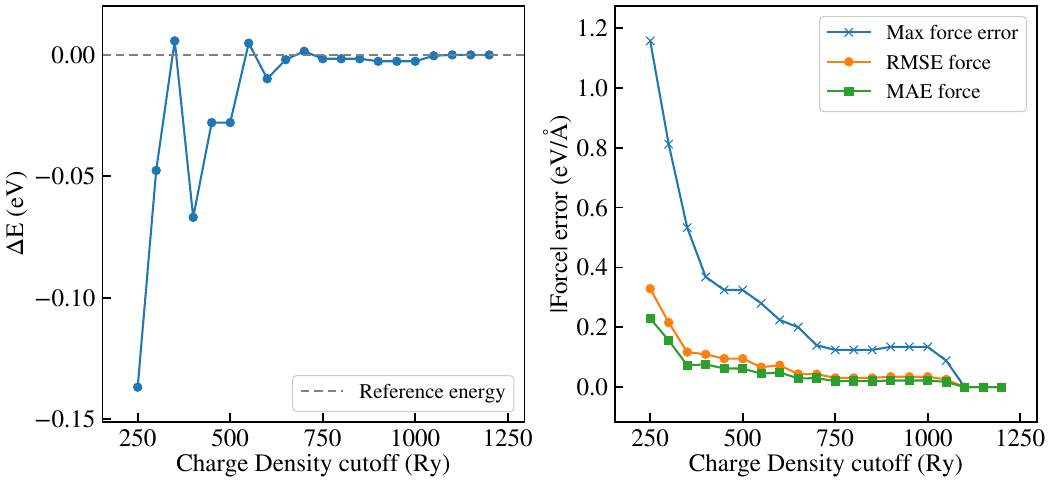}
    \caption{Convergence of the total energy (left) and force errors (right) with respect to the charge density cutoff, computed with the SCAN functional and ONCV pseudopotentials for a single amorphous In$_2$O$_3$ (a-In$_2$O$_3$) snapshot; the plane wave cutoff is held constant at 90 Ry. All errors are taken relative to the highest-cutoff calculation, which serves as the reference for both the energy and the forces. The force panel reports three measures of the per-atom force error $\Delta_i = \lvert \mathbf{F}_i - \mathbf{F}_i^{\mathrm{ref}} \rvert$: the maximum error $\max_i \Delta_i$, the root-mean-square error $\mathrm{RMSE} = \sqrt{\tfrac{1}{N}\sum_{i=1}^{N} \Delta_i^{2}}$, and the mean absolute error $\mathrm{MAE} = \tfrac{1}{N}\sum_{i=1}^{N} \Delta_i$, where the sum runs over all $N$ atoms.}
    \label{fig:placeholder}
\end{figure}

\begin{figure}[!ht]
    \centering
    \includegraphics[width=1\linewidth]{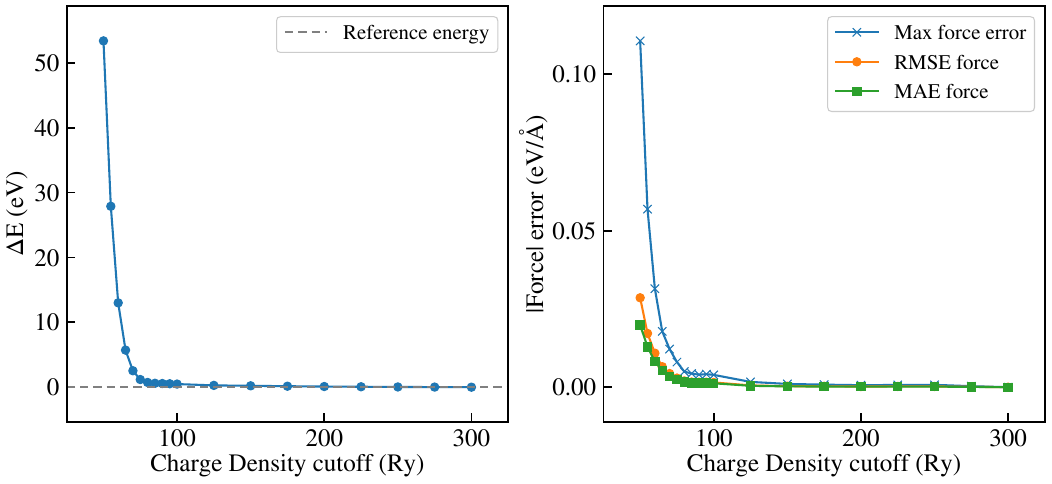}
    \caption{Convergence of the total energy (left) and force errors (right) with respect to the plane wave cutoff, computed with the SCAN functional and ONCV pseudopotentials for a single amorphous In$_2$O$_3$ (a-In$_2$O$_3$) snapshot; the charge density cutoff is held at 1200 Ry. All errors are taken relative to the highest-cutoff calculation, which serves as the reference for both the energy and the forces. The force panel reports three measures of the per-atom force error $\Delta_i = \lvert \mathbf{F}_i - \mathbf{F}_i^{\mathrm{ref}} \rvert$: the maximum error $\max_i \Delta_i$, the root-mean-square error $\mathrm{RMSE} = \sqrt{\tfrac{1}{N}\sum_{i=1}^{N} \Delta_i^{2}}$, and the mean absolute error $\mathrm{MAE} = \tfrac{1}{N}\sum_{i=1}^{N} \Delta_i$, where the sum runs over all $N$ atoms. The maximum error in forces now converges at a much lower plane wave cutoff, compared to when ecutrho is not set.}
    \label{fig:placeholder}
\end{figure}

\section{Generation of a machine learned potential for indium oxide} 
We describe below the steps followed to generate a MLIP to investigate the properties of  amorphous indium oxide.
\begin{enumerate}
    \item Using the MACE mh0 foundation model\cite{batatia2022mace, batatia2025crosslearning} and the farthest point sampling \cite{Eldar1997} (FPS) criterion we constructed an initial data set of 10,000 configurations (Set 0), drawn from 27 of 31 FPMD trajectories (80-atom samples). From the 27 trajectories we include the following configurations: 4100 structures from NVT simulations at 330 K, 5000 structures from trajectories quenched from a liquid state (3000 - 330 K), 800 structures from NVT equilibration of liquid In$_2$O$_3$ at 5000 K, and 100 structures from NVT trajectories of crystalline In$_2$O$_3$. We note that data sets to augment the foundation model are often built using random structures. Here we did not focus on designing the smallest training set but rather we built a representative one from the data already available from several indium oxide simulations that were carried out to investigate the system at the first principles level of theory. 
    \item 	Set 0 was then split 90/10 into a training (Set T1) and a validation set (Set V1). Set V1 is supplemented with data from the 4 trajectories (out of 31) that had been initially withheld from the training Set 0. Specifically, we added 20 DFT labeled configurations from each of the 4 trajectories to the validation set. 
    \item We fine-tuned the foundation model using Set T1 and used V1 for validation: we trained directly from the foundation model's weights rather than from random initialization, optimizing with the AMSGrad variant of the Adam optimizer\cite{Reddi2019}, and we used a learning rate lower than the one used for training from scratch, to avoid forgetting knowledge of the data the foundation model was trained on. We call the potential obtained at the end of this training Gen1.
    \item Using the MLIP Gen1, we carried out melt-and-quench simulations on 80 atoms cells to generate additional configurations. Specifically, we performed 50 independent melt and quench simulations (75 ps equilibration of liquid In$_2$O$_3$, followed by 25 quenches of 300 K/ps and 25 quenches of 30 K/ps), followed by a 25 ps NVT simulation. From the latter, we selected configurations for DFT labeling (e.g. configurations not well represented by the Gen 1 model) and performed SCF calculations on each flagged candidate using the same DFT parameters as for Set T1. In total we selected 1319 snapshots from all the Gen 1 simulations, and again applied a 90/10 split, yielding a new training set (Set T2) and a validation increment (Set V1.5). Set V1.5 is added to set V1 to form an updated validation set V2. Combining validation data from different sources allows the early-stopping criterion to detect both overfitting and catastrophic forgetting: because the validation set probes both the newly added data and the configurations representative of the model's prior knowledge, a rise in the validation loss flags degradation of either kind. The MACE framework computes a validation loss each epoch and stops training once this loss fails to improve for a given number of consecutive epochs (we set the number at 25).
    \item We fine-tuned the Gen1 potential on set T2 with considered V2 as the validation set.
    \item We repeated the active learning loop (steps 3-5) until no new structures were selected by the FPS criterion (i.e. we stopped the active learning procedure when every candidate from the new trajectories fell  inside the existing training distribution). Convergence required 2 additional generations, Gen 2 (540 snapshots selected) and Gen3 (with no snapshots selected, indicating convergence of the active learning procedure).
    \item[]Steps 1-6 constitute the active learning procedure carried out with 80 atom configurations.

    \item Using our final 80-atom model (Gen 3), we performed melt-and-quench simulations with 640-atom supercells. Specifically:  we carried out 50 independent melt and quench simulations (75 ps equilibration of liquid In$_2$O$_3$, followed by 25 quenches of 300 K/ps and 25 quenches of 30K/ps), followed by a 25 ps NVT simulation. From the latter trajectories, we select 78 snapshots. An additional active learning cycle confirmed that the procedure had converged with Gen3. 
    \item Once active learning had converged for both 80- and 640-atom systems, we aggregated all training and validation data into a single final pair of sets, TF and VF. Using TF and VF, we trained our final production model using the foundation model (Gen 0). Using Gen 0 mitigates any catastrophic forgetting that may have accumulated during active learning .
\end{enumerate}

\section{Validation of Machine learned Potential }
\subsection{Parity plots }

\begin{figure}[!ht]
    \centering
    \includegraphics[width=1\linewidth]{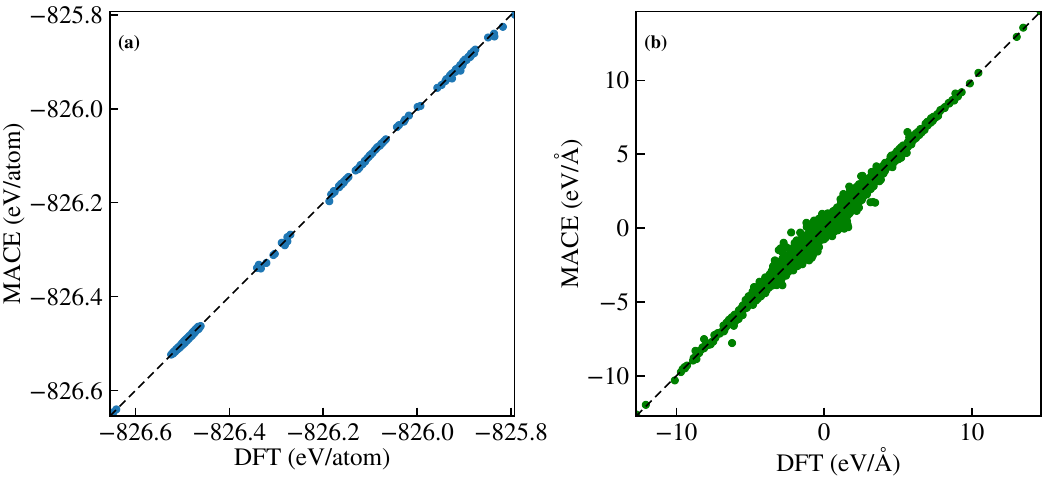}
    \caption{Parity Plots of Energy (a) and Forces (b) computed for the final validation set (VF, see text). }
    \label{fig:ML_parity_v1}
\end{figure}

In figure \ref{fig:ML_parity_v1} we show a plot of the MACE energies and forces vs the DFT energies and forces for structures in our final validation (VF) set. 

Our production model has an error of 0.003 eV/atom on energies and 0.082 eV/\AA{}  on forces for the validation set. 

\subsection{Structural Validation}

\begin{figure}[!ht]
    \centering
    \includegraphics[width=1\linewidth]{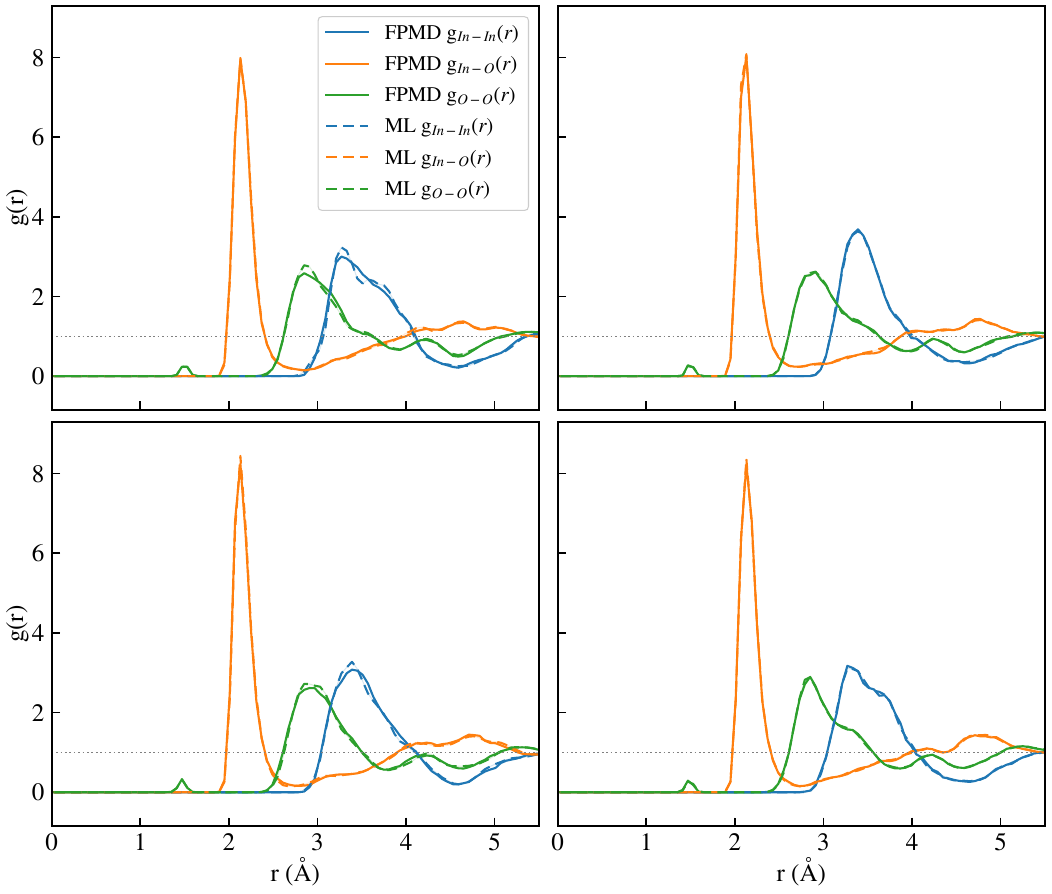}
    \caption{Partial distribution functions for 4 selected trajectories, not present in the training set, computed with MLIP and FPMD.}
    \label{fig:ml_aimd_gr}
\end{figure}

\begin{figure}[!ht]
    \centering
    \includegraphics[width=1\linewidth]{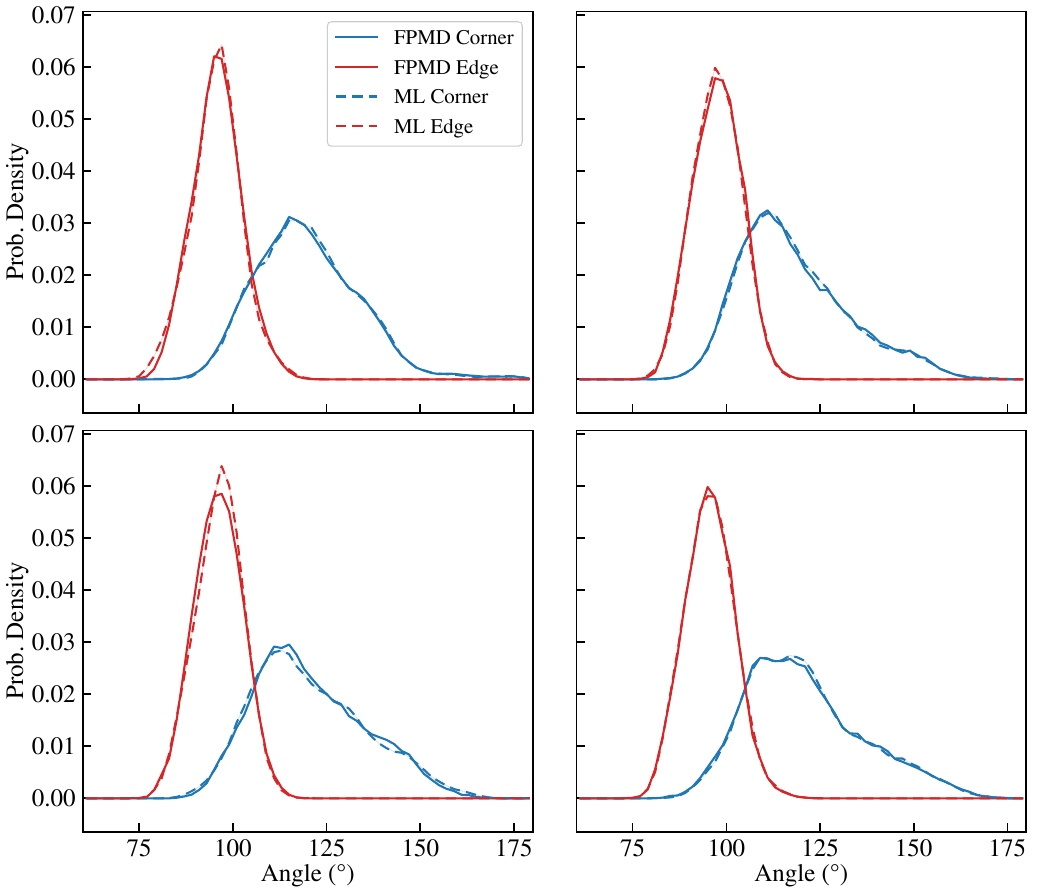}
    \caption{Angular distribution functions for 4 selected trajectories, not present in the training set, computed with MLIP and FPMD.}
    \label{fig:ml_aimd_adf}
\end{figure}

 We performed structural validations of the MLIP on 80 atom systems. Starting from the 4 FPMD generated structures not included in the training set, we carried out NVT simulations at 330 K, and compared the results for partial correlation functions and angular distributions to those of FPMD. As shown in figure \ref{fig:ml_aimd_gr}  and \ref{fig:ml_aimd_adf}, the agreement is excellent.  

\subsection{Total energies of 80 atom cells in training set computed by MACE}

Shown below (Fig. S9) is a histogram of total energies computed using the MACE model compared to the DFT energies for the same structures. 

\begin{figure}[!ht]
    \centering
    \includegraphics[width=1\linewidth]{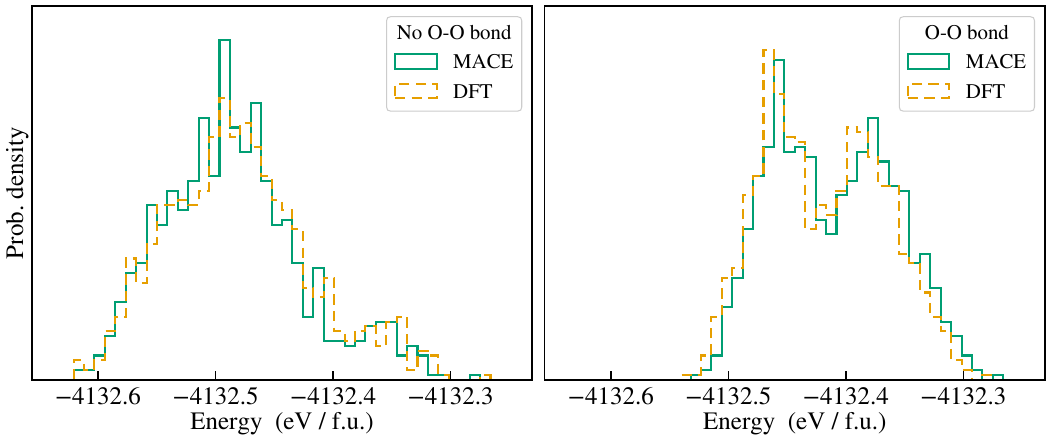}
    \caption{Distribution of total energies of samples with (left panel) and without (right panel)  OO bonds, computed with 80 atom cells for samples in the training set. We compare results obtained with the MLIP (MACE) and DFT.}
    \label{fig:mace_80_ehist}
\end{figure}

\section{Comparison of structural properties with literature results}
\begin{figure}[!ht]
    \centering
    \includegraphics[width=1\linewidth]{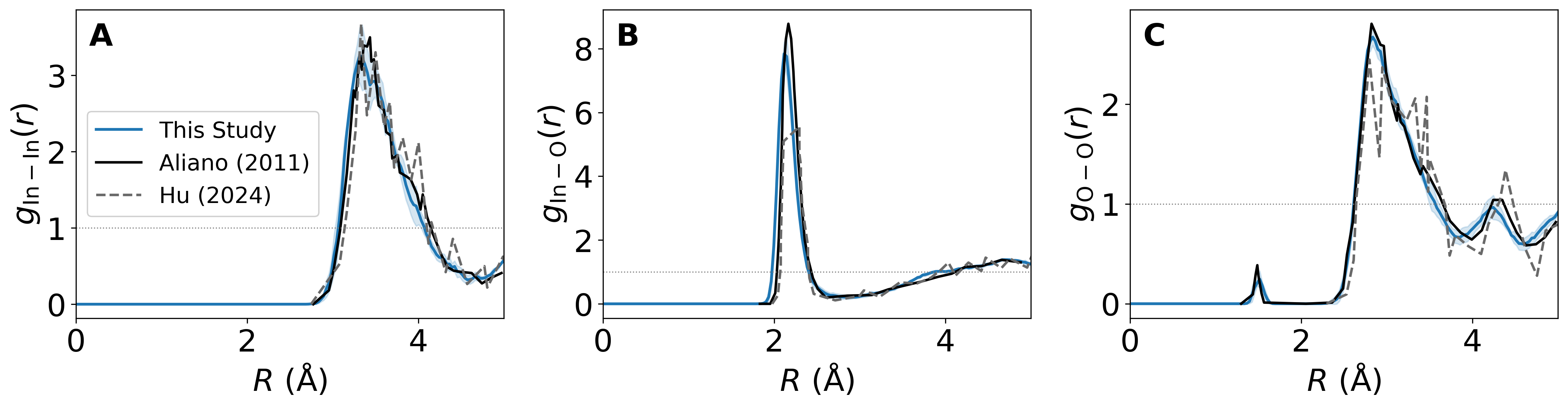}
    \caption{Partial correlation functions $g(r)$ compared with literature results from Aliano et al. \cite{Aliano2011} and Hu et al. \cite{Hu2024}. Panel A shows g$_{In-In}$(r), panel B shows g$_{In-O}$(r), and panel C shows g$_{O-O}$(r). }
    \label{fig:sio_lit_pgr}
\end{figure}

In Figure \ref{fig:sio_lit_pgr}) we compare results for partial correlation frunctions obtained with FPMD and 80 atom cells with those reported by Aliano et al. \cite{Aliano2011} and Hu et al. \cite{Hu2024}. The agreement is satisfactory. 

\begin{figure}[!ht]
    \centering
    \includegraphics[width=1\linewidth]{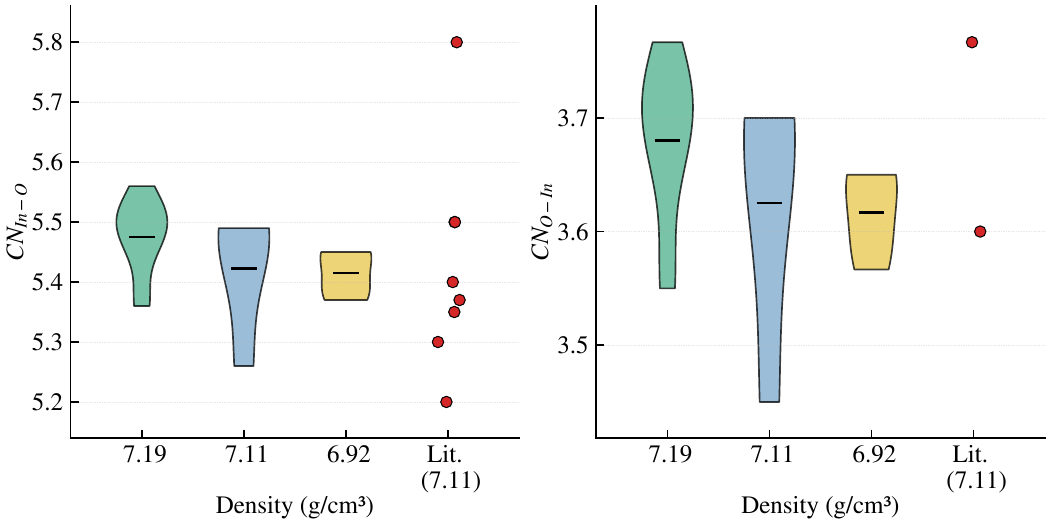}
    \caption{Violin plot comparing FPMD results with published literature results for In-O and O-In coordination numbers, defined as the number of nearest neighbors within the first coordination shell (as determined by the $g(r)$s). There are 31 samples at $\rho$ = 7.19 g $cm^{-3}$, and 4 at the other densities. Literature results taken from \cite{Aliano2011, Buchholz2014, Medvedeva2017, Khanal2015, Medvedeva2022, Medvedeva2022-2, Hu2024, Jankousky2026}}
    \label{fig:Lierature_cn}
\end{figure}

\begin{figure}
    \centering
    \includegraphics[width=1\linewidth]{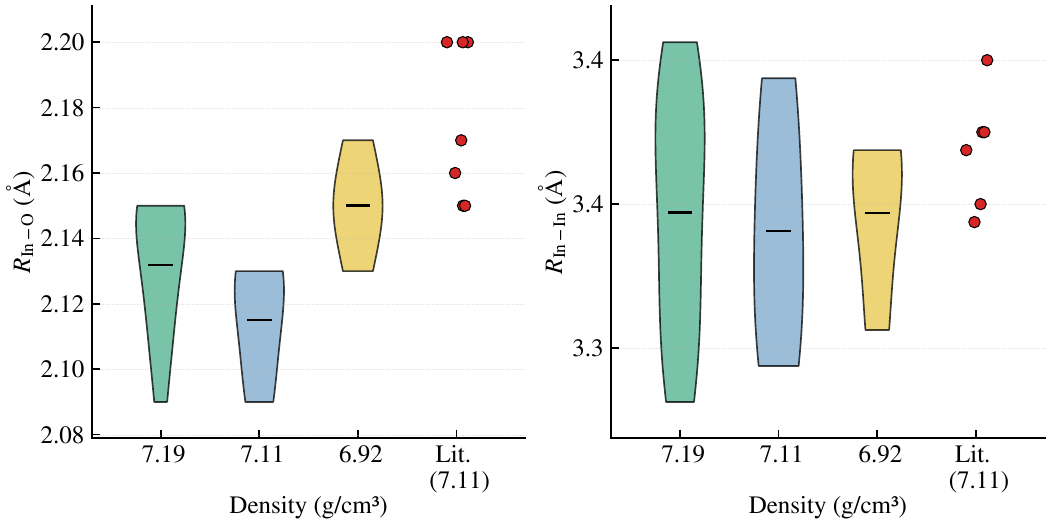}
    \caption{Violin plot comparing  FPMD results with published literature results for the In-O distance and In-In distance (as determined from  the first peak of the $g(r)$s). There are 31 samples at $\rho$ = 7.19 g $cm^{-3}$, and 4 at the other densities. Literature results taken from \cite{Aliano2011, Buchholz2014, Medvedeva2017, Khanal2015, Medvedeva2022, Medvedeva2022-2, Hu2024, Jankousky2026}}
    \label{fig:Lierature_dist}
\end{figure}

In figure \ref{fig:Lierature_cn} we compare the computed In-O and O-In coordination numbers with those reported in Ref.\cite{Aliano2011, Buchholz2014, Medvedeva2017, Khanal2015, Medvedeva2022, Medvedeva2022-2, Hu2024, Jankousky2026}. In figure \ref{fig:Lierature_dist} we compare our In-O and In-In bond distances, computed from the first peak in the $g(r)$. Distances between neighboring oxygen atoms are not reported in the literature. While our results agree well with reported In coordination numbers and In-In bond distances, our reported In-O bond distances are shorter than those observed in the literature. The results are shown in numerical form in table \ref{tab:lit-summary}

\begin{table}[H]
    \centering
    \begin{adjustbox}{max width=1.1\textwidth}
    \begin{tabular}{|c|c|c|c|c|c|c|c|}
    \hline
      Source & In-In CN & O-O CN & R$_{In-O}$ (\r{A}) & R$_{In-In}$ (\r{A}) & Method & Supercell & Density (g $cm^{-3}$) \\
      \hline
      This Study & $5.48 \pm 0.06$ & $3.65 \pm 0.04$ & $2.13 \pm 0.02$ & $3.36 \pm 0.07$ & SCAN & 80 & 7.19 \\
      This Study & $5.42 \pm 0.11$ & $3.62 \pm 0.07$ & $2.12 \pm 0.02$ & $3.35 \pm 0.07$ & SCAN & 80 & 7.11 \\
      This Study & $5.42 \pm 0.04$ & $3.61 \pm 0.02$ & $2.15 \pm 0.02$ & $3.36 \pm 0.05$ & SCAN & 80 & 6.92 \\
      \hline
      This Study &$5.47 \pm 0.02$ & $3.64 \pm 0.05$ & $2.13 \pm 0.02$  & $3.36 \pm 0.05$ & MACE & 80 & 7.19 \\
      This Study & $5.52 \pm 0.01$ & $3.68 \pm 0.06$ & $2.14 \pm 0.01$ & $3.35 \pm 0.04$ & MACE & 80 & 7.11 \\
      This Study & $5.50 \pm 0.02$ & $3.66 \pm 0.02$ & $2.13 \pm 0.02$ & $3.36 \pm 0.04$ & MACE & 640 & 7.11 \\
      This Study & $5.51$ & $3.67$ & 2.13 & 3.39 & MACE & 5120 & 7.11 \\
      \hline
      Aliano2011\cite{Aliano2011}      & 5.8  & N/A & 2.15 & 3.4  & PBE  & 105 & 7.12 \\
      Buchholtz2014\cite{Buchholz2014} & 5.2  & N/A & 2.16 & N/A  & PBE  & 80  & 7.12 \\
      Khanal2015\cite{Khanal2015}      & 5.4  & N/A & 2.17 & 3.35 & PBE  & 130 & 7.12 \\
      Medvedeva2017\cite{Medvedeva2017}& 5.3  & N/A & 2.20 & 3.4  & PBE  & 80  & 7.12 \\
      Allec2020\cite{Allec2020}        & 5.37 & 3.6 & 2.20 & 3.36 & PBE  & 80  & 7.12 \\
      Medvedeva2022\cite{Medvedeva2022}& 5.35 & N/A & 2.20 & 3.39 & PBE  & 135 & 7.12 \\
      Hu2024\cite{Hu2024}              & 5.5  & 3.7 & N/A  & 3.44 & PBE  & 80 & 7.12 \\
      Jankousky2026 \cite{Jankousky2026} & 5.3  & N/A & N/A  & 3.6  & PBE & 40   & 7.12 \\
      \hline
    \end{tabular}
    \end{adjustbox}
    \caption{Summary of  structural properties for a-In$_2$O$_3$, compared with published results\cite{Aliano2011, Buchholz2014, Medvedeva2017, Khanal2015, Medvedeva2022, Medvedeva2022-2, Hu2024, Jankousky2026}. Coordination numbers and bond distances determined from the first peak of the $g(r)$s are also shown in Fig. S11 and S12, respectively.}
    \label{tab:lit-summary}
\end{table}

\section{Computed partial correlation functions with error bars}
\label{section:gr_error}

This section shows the computed partial $g(r)$ with error bars for our FPMD generated and MLIP generated samples. 

\begin{figure}[!ht]
    \centering
    \includegraphics[width=1\linewidth]{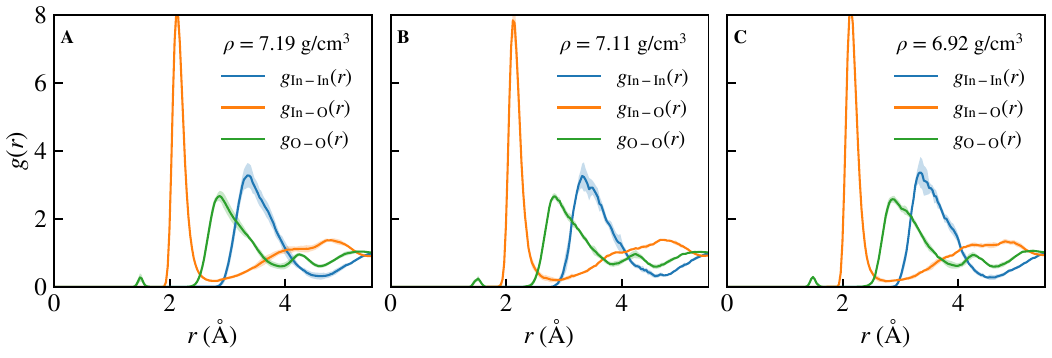}
    \caption{Partial radial distribution functions, g$_{In-In}$(r) (left), g$_{In-O}$(r) (center), and g$_{O-O}$(r) (right), computed for 80 atom systems generated with FPMD as a function of density, with error bars shown. For the density of 7.19 g $cm^{-3}$, 23 configurations are averaged, for the remaining densities 4 configuration are averaged.  }
    \label{fig:si_pgr_rho}
\end{figure}

\begin{figure}[!ht]
    \centering
    \includegraphics[width=0.5\linewidth]{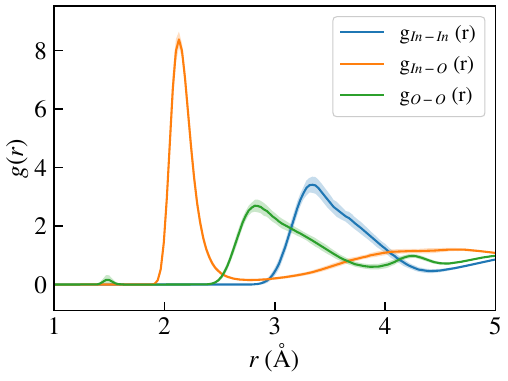}
    \caption{Partial distribution functions for 80 atom cells with a density of $\rho$ = 7.19 g$\cdot$ cm$^{-3}$, averaged across 50 separate configurations.}
    \label{fig:placeholder}
\end{figure}

\begin{figure}[!ht]
    \centering
    \includegraphics[width=0.5\linewidth]{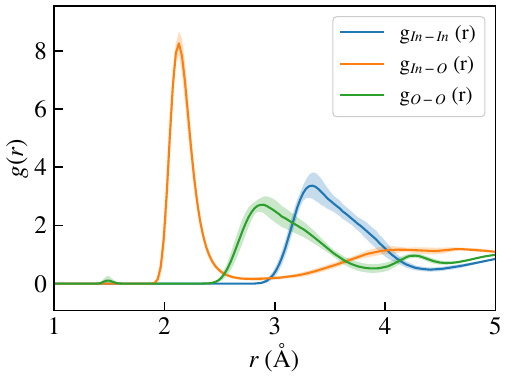}
    \caption{Partial distribution functions for 80 atom cells with a density of $\rho$ = 7.11 g$\cdot$ cm$^{-3}$, averaged across 45 separate configurations.}
    \label{fig:placeholder}
\end{figure}

\begin{figure}[!ht]
    \centering
    \includegraphics[width=.5\linewidth]{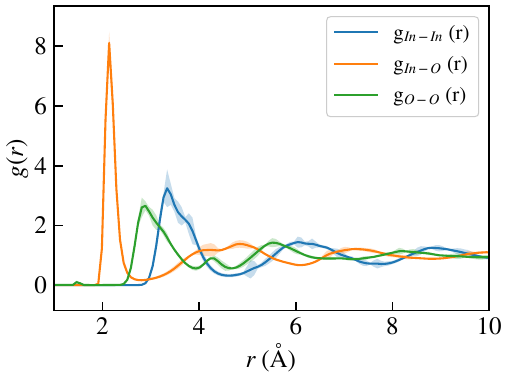}
    \caption{Partial distribution functions for 640 atom supercells at $\rho$ = 7.11 g$\cdot$ cm$^{-3}$, averaged across 45 separate configurations. }
    \label{fig:si_pgr_640}
\end{figure}

\begin{figure}[!ht]
    \centering
    \includegraphics[width=.5\linewidth]{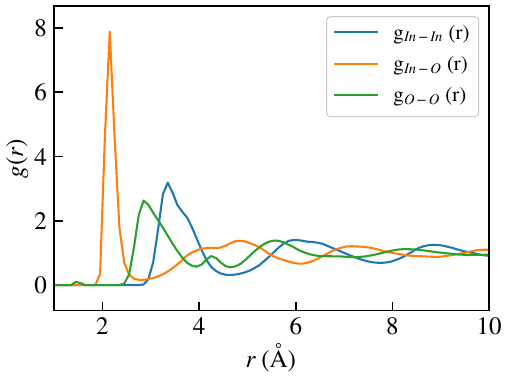}
    \caption{Partial distribution functions for a 5120 atom cell at $\rho$ = 7.11 g$\cdot$ cm$^{-3}$}
    \label{fig:pgr_5120}
\end{figure}





\section{Computed interference functions for different cell sizes}

\begin{figure}[!ht]
    \centering
    \includegraphics[width=1\linewidth]{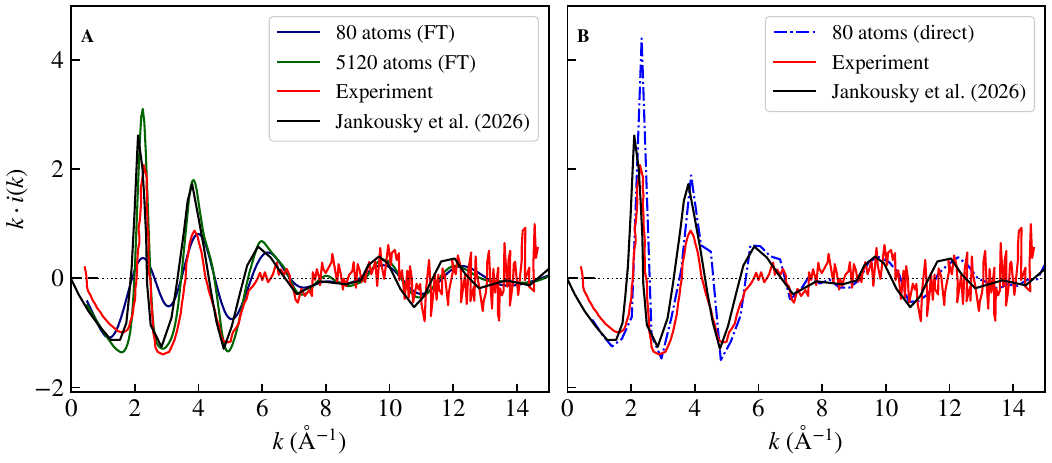}
    \caption{Panel A: Comparison of k $\cdot$ i(k) computed by Fourier transforming the total radial distribution function obtained from FPMD in a  80 atom  cell and with MLIP in a 5120 atom cell;  experimental results are also shown\cite{Utsuno2006} Panel B: Comparison of k $\cdot$ i(k) computed in Fourier space from FPMD for a 80 atom cell with experiment and the calculations of Ref \cite{Jankousky2026}, which averages over 1500 samples.}
    \label{fig:sk_lit}
\end{figure}

Figure \ref{fig:sk_lit}A shows the interference function k $\cdot$ i(k) computed using equation \ref{eq:ik_gr} of the main text, for  80 atom and 5120 atom systems. The effect of finite size can be seen in the first sharp diffraction peak, which is much smaller and broader in the 80 atom systems. We also computed the total reduced X-ray interference function directly in reciprocal space, using equation \ref{eq:ik_direct} in the main text. Figure \ref{fig:sk_lit}B shows  our results for 80 atom systems compared with the results by Jankousky et al. \cite{Jankousky2026}. Note that Jankousky et al. averaged over 1500 samples. 

\section{Calculation of IR spectra}

\begin{figure}[!ht]
    \centering
    \includegraphics[width=0.5\linewidth]{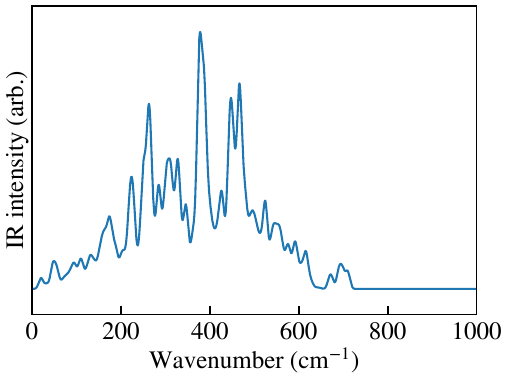}
    \caption{IR spectrum for 80 atom system containing an OO bond. Notably, the mode at $\sim$ 850 cm$^{-1}$ is not IR active. }
    \label{fig:aIO_ir}
\end{figure}

Figure \ref{fig:aIO_ir} shows the IR spectra for our 80 atom system with an O-O bond, computed at the LDA level. 

\section{Angular Distribution Functions for samples generated by FPMD }

\begin{figure}[!ht]
  \centering
  \includegraphics[width=1\linewidth]{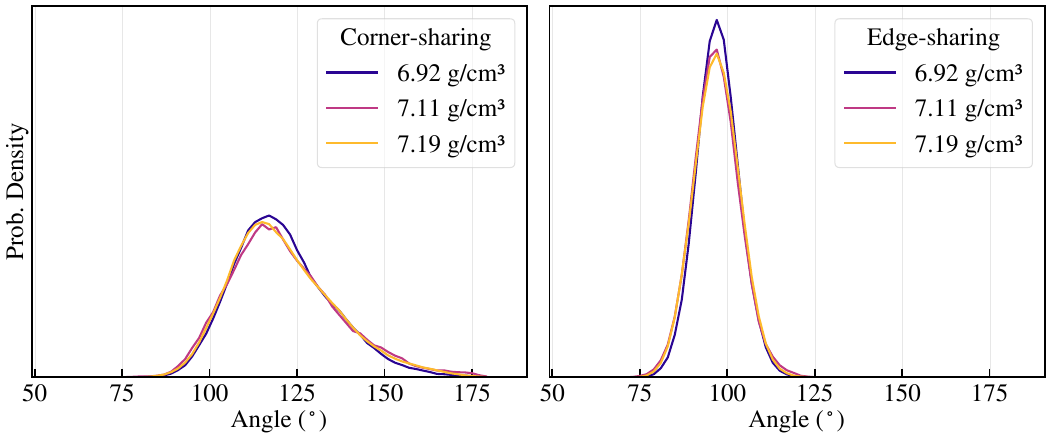}
  \caption{In--O--In bond-angle distributions in amorphous In$_2$O$_3$ as a function of mass density. (a) Corner-sharing configurations, in which two In$_2$O$_3$ polyhedra share a single bridging oxygen, and (b) Edge-sharing configurations, in which two polyhedra share two bridging oxygens.}
  \label{fig:aIO_adf_rho}
\end{figure}

Figure \ref{fig:aIO_adf_rho} shows the computed angular distribution function for FPMD generated samples at different densities. Little variation is found across the densities studied.  

\section{Angular Distribution Functions with error bars }
\label{section:adf_error}
This section shows the computed partial angular distribution functions, broken down by polyhedra type, with error bars for FPMD generated and MLIP generated.

\begin{figure}[!ht]
   \centering
    \includegraphics[width=1\linewidth]{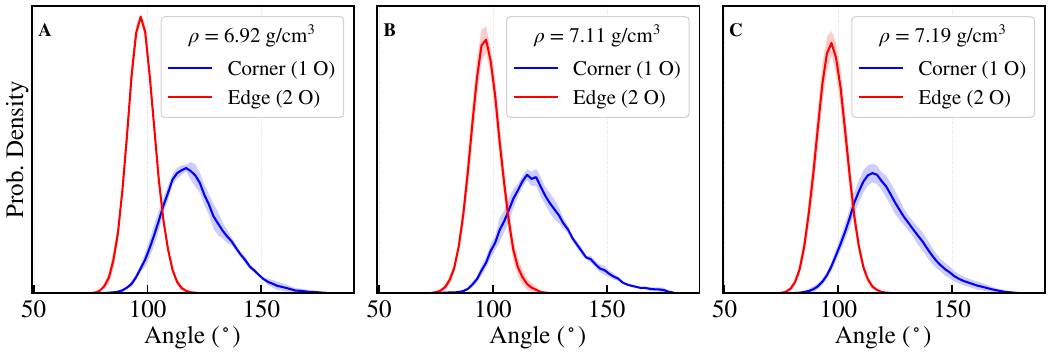}
    \caption{Angular distribution functions computed for 80 atom systems as a function of density with error bars. For the denisty of 7.19 g$\cdot$ cm$^{-3}$, 23 configurations are averaged, for the remaining densities 4 configuration are averaged.}
    \label{fig:adf_error}
\end{figure}

\begin{figure}
    \centering
    \includegraphics[width=0.5\linewidth]{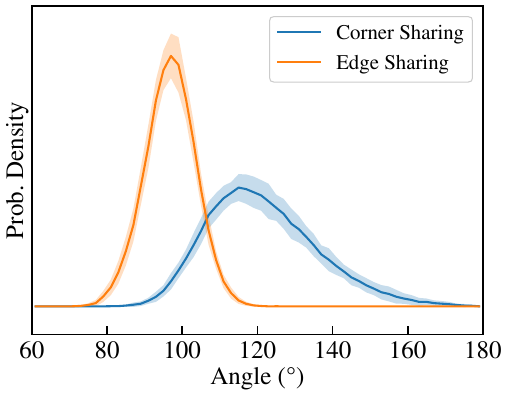}
    \caption{Angular distribution functions computed for 80 atom systems at a density of 7.19 g$\cdot$ cm$^{-3}$, averaged for 50 configurations with error bars shown}
    \label{fig:adf_640}
\end{figure}

\begin{figure}
    \centering
    \includegraphics[width=0.5\linewidth]{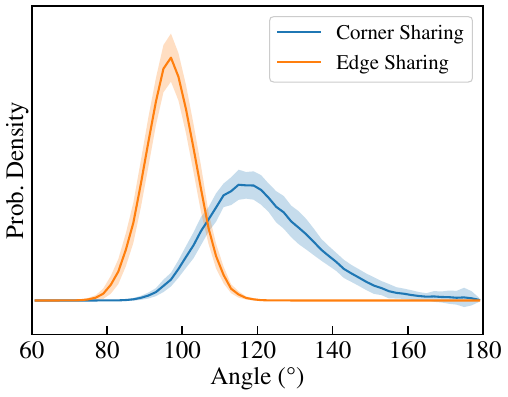}
    \caption{Angular distribution functions computed for 80 atom systems at a density of 7.11 g$\cdot$ cm$^{-3}$, averaged for 45 configurations with error bars shown}
    \label{fig:adf_640}
\end{figure}

\begin{figure}[!ht]
    \centering
    \includegraphics[width=0.5\linewidth]{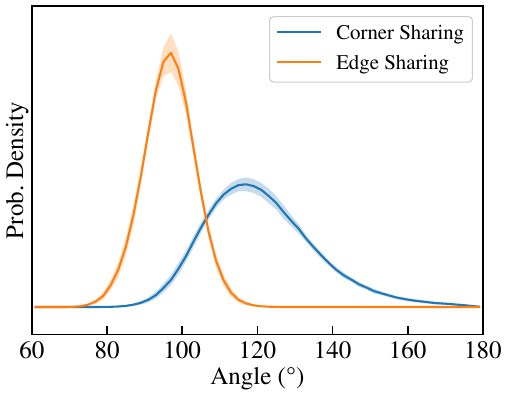}
    \caption{Angular distribution functions computed for 640 atom systems at a density of 7.11 g$\cdot$ cm$^{-3}$, averaged for 45 configurations with error bars shown}
    \label{fig:adf_5120}
\end{figure}

\begin{figure}[!ht]
    \centering
    \includegraphics[width=0.5\linewidth]{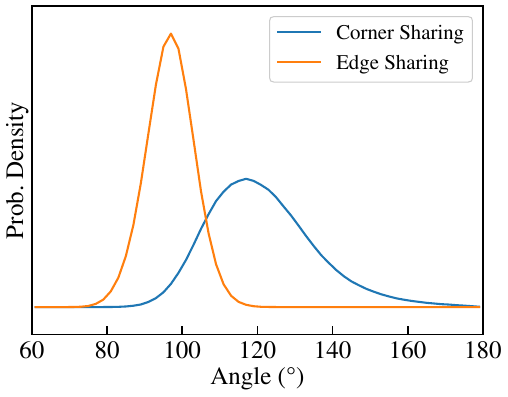}
    \caption{Angular distribution functions computed for a 5120 atom system at a density of 7.11 g$\cdot$ cm$^{-3}$.}
    \label{fig:placeholder}
\end{figure}

\section{Distribution of polyhedra}
\begin{figure}[!ht]
    \centering
    \includegraphics[width=0.5\linewidth]{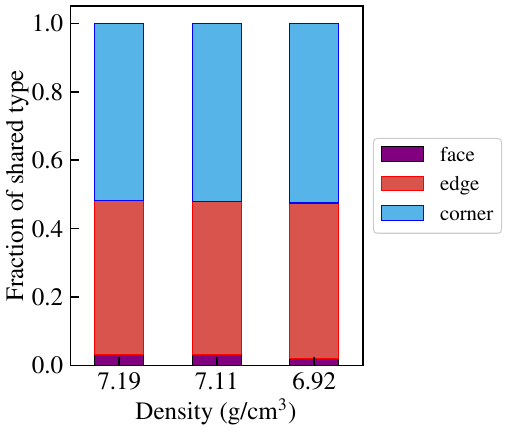}
    \caption{The percentage of edge, face and corner  sharing polyhedra as a function of density}
    \label{fig:poly_frac}
\end{figure}

\begin{figure}[!ht]
    \centering
    \includegraphics[width=1\linewidth]{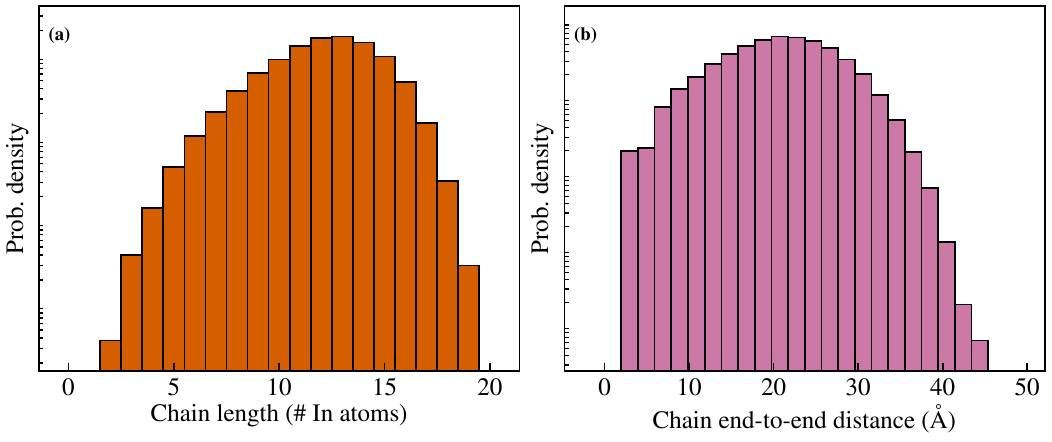}
    \caption{Histogram of Chain lengths as a function of the number of atoms (a) and length (b) for a 5120 atom supercell with density of 7.11 g cm$^{-3}$.}
    \label{fig:5120_chain_stats}
\end{figure}

\begin{figure}[!ht]
    \centering
    \includegraphics[width=1\linewidth]{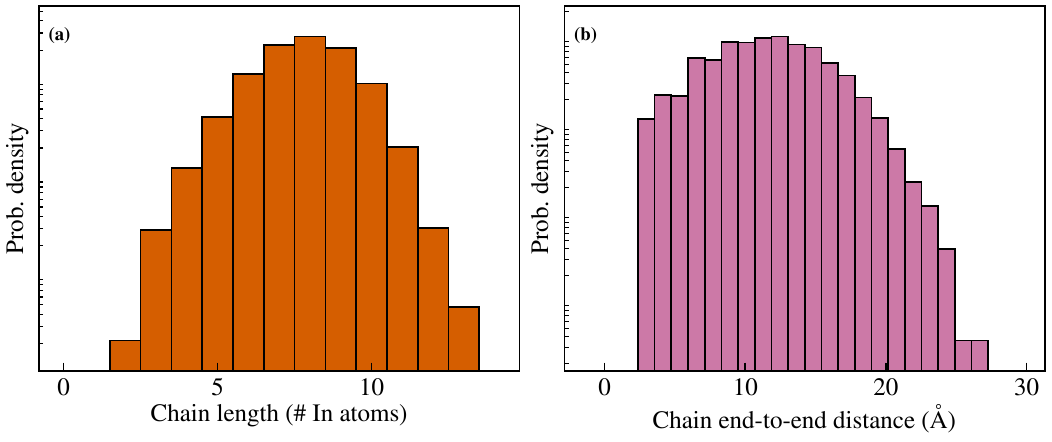}
    \caption{Histogram of Chain lengths as a function of the number of atoms (a) and length (b) for a 5120 atom supercell with density of 7.11 g cm$^{-3}$.}
    \label{fig:chain_stats}
\end{figure}

\begin{figure}[!ht]
    \centering
    \includegraphics[width=0.5\linewidth]{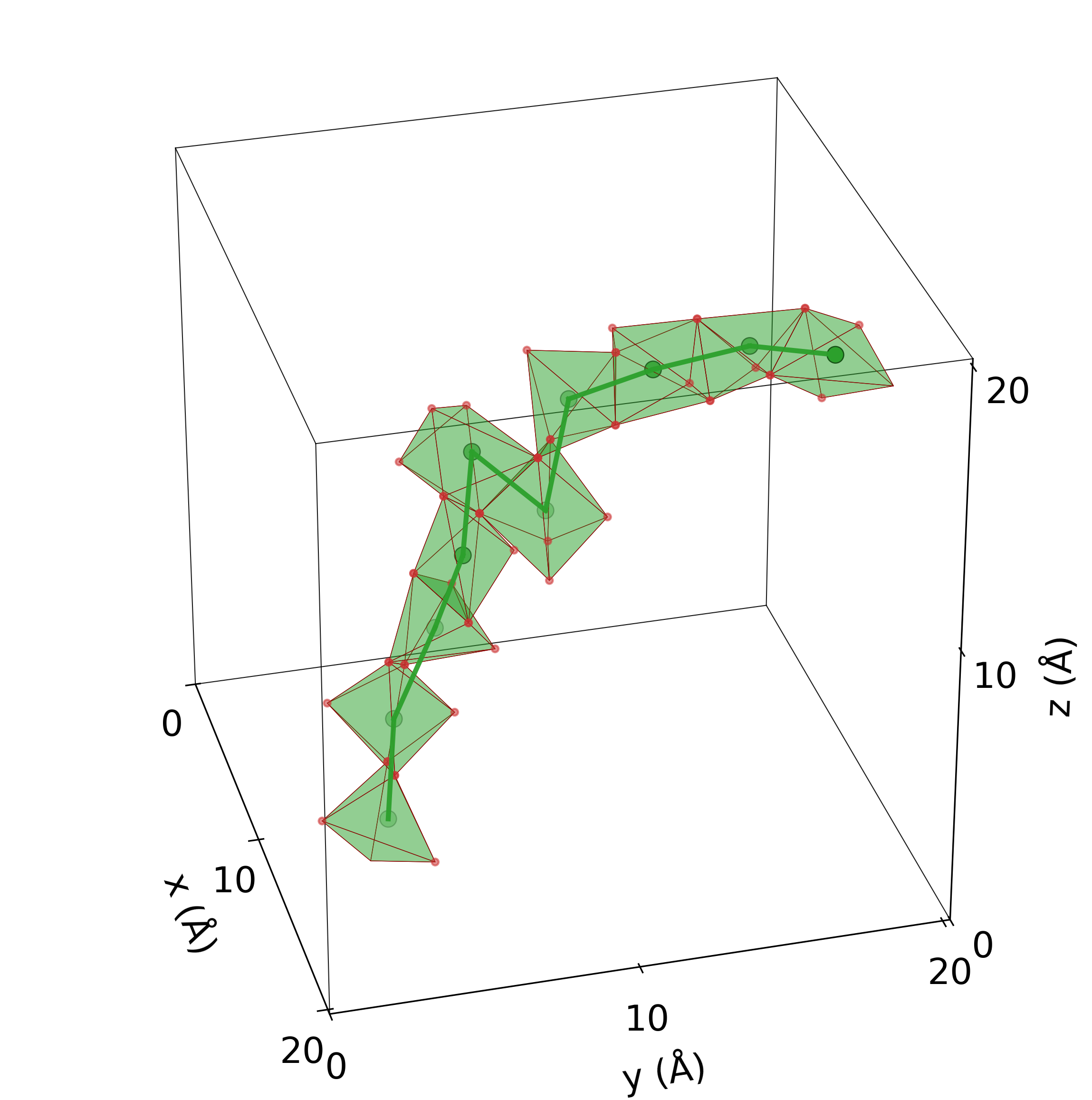}
    \caption{Longest chain of InO edge sharing polyhedra, shown for a single 640 atom supercell with density of 7.11 g cm$^{-3}$.} 
    \label{fig:chain}
\end{figure}

Figure \ref{fig:poly_frac} shows the percentage of edge, face and corner sharing polyhedra, defined as sharing 1, 2, or 3 oxygens, as a function of density. Figures \ref{fig:5120_chain_stats} and \ref{fig:chain_stats} show the distribution of chain lengths in the edge connected polyhedral network for 5120 and 640 atom supercells, while Figure \ref{fig:chain} shows the longest chain in our 640 atom supercells.

For a 640 atom cell (cubic with length 20.24 \AA{}) we find that the average chain length is 7.82 $\pm$ 1.46 In atoms, spanning 11.54 $\pm$ 3.81 \AA{}. Some of the chains extend across the entire cell as shown in figure \ref{fig:chain}, functioning as pathways for conduction electrons.

\section{Diffusion of atoms in Liquid In$_2$O$_3$}

\begin{figure}[!ht]
    \centering
    \includegraphics[width=1\linewidth]{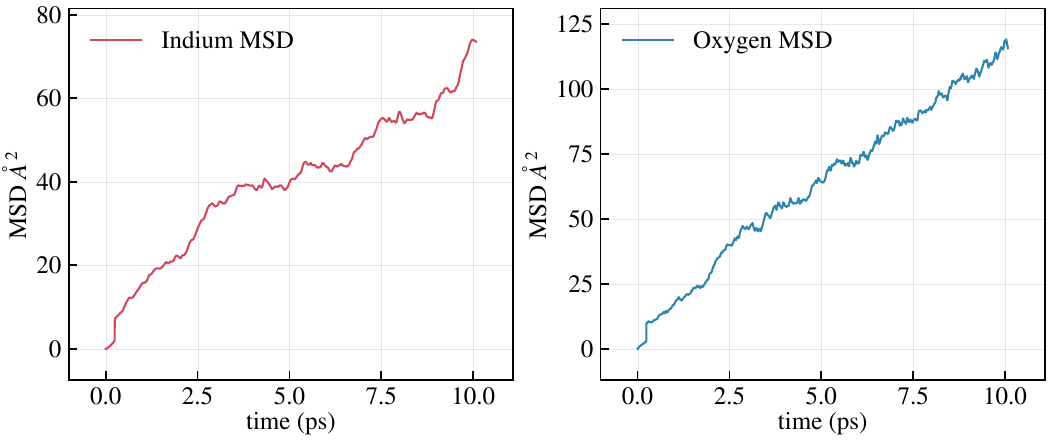}
    \caption{Mean squared displacement (MSD) of indium (left) and oxygen (right) atoms in 80 FPMD as a function of time during equilibration at 5000 K.}
    \label{fig:80_liquid}
\end{figure}

\begin{figure}[!ht]
    \centering
    \includegraphics[width=1\linewidth]{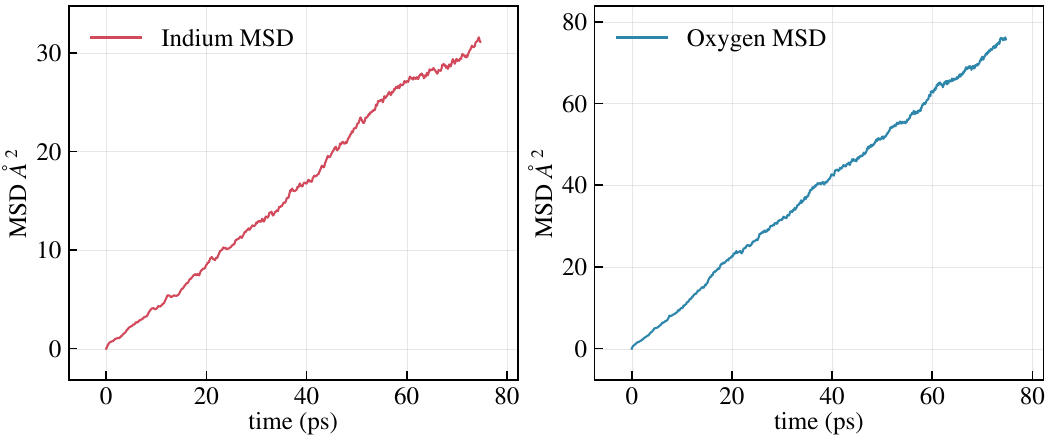}
    \caption{Mean squared displacement (MSD) of indium (left) and oxygen (right) atoms in 640 atom supercells as a function of time during equilibration at 3000 K.}
    \label{fig:640_liquid}
\end{figure}

We show below the mean squared displacement (MSD) of indium and oxygen atoms in representative supercells as a function of time during equilibration runs in the liquid state. Figure \ref{fig:80_liquid} shows diffusion during the equilibration of our liquid sample during FPMD, at 5000 K, while \ref{fig:640_liquid} shows the diffusion during the equilibration of one of our 640 melt and quenches.

\section{Histogram of total energies by OO bond count in 80 atom systems as a function of density}

\begin{figure}[!ht]
    \centering
    \includegraphics[width=1\linewidth]{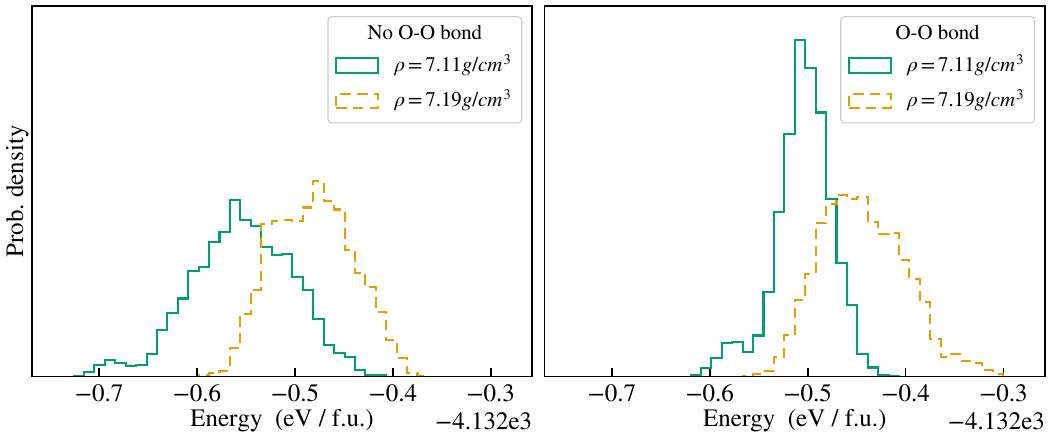}
    \caption{Energies per formula unit computed with the MLIP for 80 atom samples collected at two densities, separated by the presence of an O-O bond in the system.}
    \label{fig:80v80}
\end{figure}

Figure \ref{fig:80v80} shows the distribution of MACE energies per formula unit computed with the MLIP for 80 atom samples collected at two densities. The samples at 7.11 g $cm^{-3}$ are found at lower energies. 

\bibliography{bib-am}